\documentclass[9pt]{ieeeconf}

\usepackage{amsmath,amssymb,amsfonts}
\usepackage{bm}
\usepackage{graphicx}
\usepackage{xcolor}
\usepackage{textcomp}
\usepackage{pifont}
\usepackage{tabularx}
\usepackage{booktabs}
\usepackage{multirow}
\usepackage{threeparttable}
\usepackage{float}
\usepackage{algorithm}
\usepackage{algpseudocode}
\usepackage{enumerate}
\usepackage{diagbox}
\usepackage{placeins}
\usepackage{amsthm}
\usepackage{comment}
\usepackage{etoolbox}
\usepackage{xspace}
\usepackage{hyperref}
\usepackage[capitalize,nameinlink]{cleveref}
\crefname{equation}{}{}
\Crefname{equation}{Equation}{Equations}

\theoremstyle{plain}

\newtheorem{definition}{Definition}

\renewcommand{\algorithmicrequire}{\textbf{Input:}}
\renewcommand{\algorithmicensure}{\textbf{Output:}}
\renewcommand{\algorithmiccomment}[1]{\bgroup\hfill$\triangleright$~#1\egroup}
\makeatletter
\def\theHALG@line{\thealgorithm.\arabic{ALG@line}}
\makeatother

\usepackage{caption}
\def\tsc#1{\csdef{#1}{\textsc{\lowercase{#1}}\xspace}}
\tsc{WGM}
\tsc{QE}
\tsc{EP}
\tsc{PMS}
\tsc{BEC}
\tsc{DE}

\title{Poisoning Attacks to Local Differential Privacy Protocols for \\Dynamic Graphs}

\addauthor{Shang Liu\thanks{Our code is available at: \url{https://github.com/spiralfood/GraphPoison-LDP}.}}{1,2}
\addauthor{Yuhan Bao}{1}
\addauthor{Hongbin Zhao}{1}
\addauthor{Yanmei Zhang}{1}
\addauthor{Guan Yuan}{1}
\addauthor{Yang Cao}{3}

\addaffiliation{1}{School of Computer Science and Technology/School of Artificial Intelligence, China University of Mining and Technology
}
\affiliationbreak
\addaffiliation{2}{Mine Digitization Engineering Research Center of the Ministry of Education}
\affiliationbreak
\affiliationbreak
\addaffiliation{3}{Department of Computer Science, Institute of Science Tokyo\\
    \texttt{%
        \{shang,baoyuhan,08233112,ymzhang,yuanguan\}@cumt.edu.cn, cao@c.titech.ac.jp
    }%
}

\begin{document}
\maketitle
\pagestyle{plain}        
\pagenumbering{arabic}   
\begin{abstract}
With the increasing adoption of privacy-preserving techniques in dynamic graph analysis, local differential privacy (LDP) has become a fundamental paradigm for protecting user data in decentralized environments. However, despite its strong privacy guarantees, LDP-based dynamic graph protocols remain vulnerable to data poisoning attacks through malicious data injection. In this paper, we present the first systematic study of data poisoning attacks against privacy-preserving dynamic graphs under the LDP framework. Considering the temporal evolution of dynamic graphs, we identify key challenges in attack design and evaluation, including how to schedule attack resources across time and how to measure cumulative attack impact. To address these challenges, we introduce an absolute gain metric to quantify the overall deviation of graph statistical metrics before and after attacks across time. Based on this metric, we further propose a unified poisoning attack framework for dynamic graphs, which consists of target node selection, dynamic fake node scale scheduling, and fake edge injection. Within this framework, we design a temporal enhanced attack (TEA) strategy that dynamically schedules attack resources to maximize attack effectiveness, along with two baseline strategies, equal-scale attack (ESA) and random-scale attack (RSA), for comparison. Extensive experiments on real-world and synthetic datasets demonstrate that the proposed attacks can significantly degrade the accuracy of graph metric estimation for target nodes. 

\end{abstract}

\noindent\textbf{Keywords:} Local differential privacy; Poisoning attacks; Dynamic graphs
\par

\section{Introduction}

Dynamic graphs, whose topology evolves over time, are widely observed in real-world systems such as social networks \cite{sekara2016fundamental}, traffic networks \cite{han2014continuum}, and epidemiological networks \cite{keeling2005networks}. 
Analyzing these evolving structures enables a variety of real-time and high-impact applications, including hotspot tracking \cite{yu2013anomalous}, behavioral analysis \cite{rossi2013modeling}, and personalized advertising \cite{yin2023next}. 
For example, temporal patterns of user interactions can be used to predict how events propagate across online platforms \cite{khafaei2019tracing}. 
However, dynamic graphs often contain highly sensitive personal information, such as social relationships or purchase histories, raising significant privacy concerns and limiting their safe sharing.

Local differential privacy (LDP) \cite{lin2022towards,ye2020lf,xu2025continuous} has emerged as a promising paradigm for decentralized graph data protection. 
Unlike traditional approaches (central differential privacy, CDP \cite{dwork2014algorithmic}) that rely on a trusted curator, LDP allows each individual user to locally perturb their data before sharing it, ensuring that sensitive details are protected at the point of data collection. 
This model shifts control of privacy to the user side and eliminates the reliance on a central trusted authority. 
LDP has already been applied successfully in graph analysis such as degree distribution~\cite{day2016publishing,liu2022crypto}, subgraph counting \cite{imola2021locally,liu2024cargo}, and graph publication \cite{qin2017generating,liu2025pgb}, enabling privacy-preserving data analysis and collection without compromising user confidentiality.

Despite its strong privacy guarantees, data poisoning attacks have recently emerged as a significant threat to LDP protocols.
Prior studies have shown that LDP aggregation mechanisms can be misled by manipulated user reports~\cite{huang2024ldpguard,li2025mitigating,zhan2025poisoning}.
For instance, an adversary may inject a small number of malicious users or submit carefully crafted fake reports, thereby altering the distribution of the collected statistics and significantly biasing the final aggregated estimations.
Existing poisoning attacks against LDP protocols mainly focus on tabular data \cite{huang2024ldpguard,li2025mitigating} or static graphs~\cite{he2025data}, including tasks such as frequency estimation, heavy-hitter detection, and mean estimation.
However, these studies are largely limited to static settings, where the underlying data structure does not evolve over time.

\newcommand{\attackfig}[1]{\includegraphics[width=\linewidth,height=0.75\linewidth]{#1}}

\begin{figure*}[t]
    \centering
    \includegraphics[width=\textwidth]{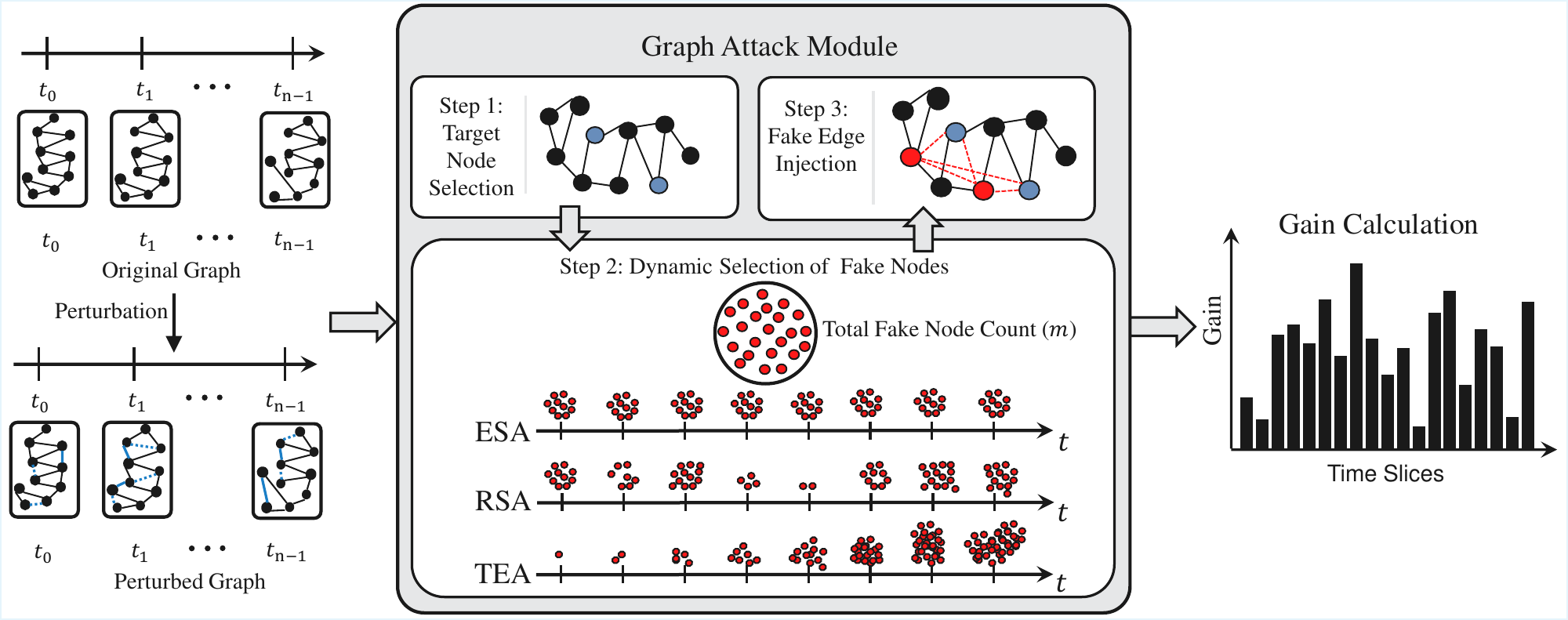}
    \caption{The proposed attack framework to LDP for dynamic graphs, consisting of (i) target node selection, (ii) dynamic selection of fake nodes, and (iii) fake edge injection.}
    \label{fig:attack_framework}
\end{figure*}

In contrast, the vulnerability of dynamic graphs under LDP to poisoning attacks remains largely unexplored. 
Compared with static graphs, dynamic graphs \cite{yin2023next,hou2023ppdu,li2025dynamic} consist of a sequence of evolving network snapshots, which introduces additional complexities for adversarial manipulation. 
Specifically, two key challenges arise in dynamic graph settings:
(1) Attackers need to determine how to evaluate the effectiveness of poisoning attacks in dynamic graphs: the influence of attacks may accumulate across multiple snapshots over time, making it difficult to directly measure the overall attack gain.
(2) Attackers have to determine how to allocate malicious actions across multiple time steps to maximize the overall attack impact: overly concentrated attacks in a single snapshot may produce strong short-term effects but increase the risk of detection, whereas overly dispersed attacks may significantly weaken the overall effectiveness under limited attack resources.

In this paper, we explore the data poisoning attacks to local differential privacy protocols for dynamic graphs for the first time.
To solve the challenge (1), we design an absolute gain metric for dynamic graph attacks to quantify the changes in graph statistical metrics before and after the attack. 
To handle the challenge (2), based on this metric, we design a temporal enhanced attack (TEA) strategy that schedules attack resources across time steps to maximize the overall attack gain in dynamic graphs. 
In addition, we construct two baseline attack strategies, namely equal-scale attack (ESA) and random-scale Attack (RSA), for comparative evaluation.

Building upon these designs, we propose a unified poisoning attack framework for LDP-based dynamic graphs, as illustrated in Fig.~\ref{fig:attack_framework}. The framework consists of three key components: (i) target node selection, which randomly selects target nodes from the final graph snapshot according to a predefined target node ratio $\gamma$; (ii) dynamic fake node scheduling, which determines the total number of fake nodes and allocates them across different time slices using three scheduling strategies, namely ESA, RSA, and TEA; and (iii) fake edge injection, which constructs and injects forged edges through three topology attack strategies, namely RVA, RNA, and MGA, to manipulate the graph metrics of target nodes. Together, these components form a complete pipeline for effective adversarial manipulation under privacy constraints.

Our main contributions are summarized as follows:

\begin{itemize}

\item We conduct the first systematic investigation of data poisoning attacks against LDP protocols in dynamic graph settings, revealing previously unexplored security vulnerabilities.
\item We propose a general poisoning attack framework tailored to dynamic graphs under LDP, along with three attack strategies (TEA, ESA, and RSA) designed to effectively manipulate graph statistics over time.
\item Extensive experiments on both real-world and synthetic datasets demonstrate that our proposed attacks can significantly degrade the accuracy of graph metric estimation for target nodes, highlighting the urgent need for robust defense mechanisms in privacy-preserving dynamic graph systems.
\end{itemize}

The remainder of this paper is organized as follows. Section \ref{sec:related work} reviews the existing literature on privacy-preserving graph analysis and data poisoning attacks, while Section \ref{sec:preliminary} introduces the necessary background and problem formulation. Section \ref{sec:method} presents the proposed data poisoning attack framework, including the design of three attack strategies and the corresponding gain metric for dynamic graphs. 
Section \ref{sec:evaluation} provides a comprehensive experimental evaluation on both real-world and synthetic datasets, analyzing the effectiveness and robustness of the proposed methods. Finally, Section \ref{sec:conclusion} concludes the paper and discusses potential directions for future work.

\section{Related Work}
\label{sec:related work}
\subsection{Local Differential Privacy Protocols}
Differential Privacy (DP), as a foundational framework for privacy protection, relies on a trusted third party to collect data and inject noise, thereby strictly limiting the influence of any individual record on the output  \cite{dwork2006calibrating,dwork2014algorithmic,li2017differential}. However, in real-world applications, third-party data aggregation is often untrustworthy. The introduction of Local Differential Privacy (LDP) effectively addresses this issue by allowing users to add noise directly at the data generation stage, eliminating the dependence on a trusted third party at the mechanism level \cite{erlingsson2014rappor}.

LDP has been widely studied for protecting sensitive information in graph analysis \cite{ye2020lf,imola2021locally,9835478, wei2020asgldp,imola2022communication}. 
For instance, LF-GDPR \cite{ye2020lf} provides a general framework for graph metric estimation by locally perturbing adjacency bit vectors and node degrees, supporting tasks such as clustering coefficient estimation and community detection.
Imola et al. \cite{imola2021locally} develop LDP algorithms for triangle and \(k\)-star counting, demonstrating that interaction can improve triangle-counting accuracy.
Imola et al. \cite{imola2022communication} combine edge sampling and double clipping to reduce communication costs and estimation error in triangle counting under LDP.
To account for correlations across users' local graph views, Edge-RLDP \cite{9835478} introduces a correlation-aware privacy definition and a two-phase framework for collecting triangle counts. 
AsgLDP \cite{wei2020asgldp} collects decentralized attributed graphs under LDP and generates synthetic graphs that preserve structural and attribute statistics.

\subsection{Dynamic Graph Analysis under LDP}
Under the LDP framework, recent studies have begun to explore decentralized graph data publishing in dynamic settings. 
Such dynamic graphs are typically represented as a sequence of static graph snapshots over time. By continuously applying local perturbations across consecutive time slices, these methods achieve privacy protection without relying on a trusted central server.
For instance, Hou et al.~\cite{hou2023ppdu} constructed a node-level privacy scope and perform LDP perturbation within this scope, updating only the affected nodes when the graph evolves, thereby reducing noise and improving efficiency. 
Yuan et al.~\cite{yuan2024psgraph} adaptively allocated the privacy budget by analyzing structural changes between adjacent snapshots and generate synthetic graphs by perturbing node degrees and inter-community connections at the community level. 
Similarly, Hao et al.~\cite{hao2025caldp} exploited community structures to mitigate perturbation impact by restricting perturbation ranges, applying differentiated strategies for edges, and updating only the changed communities to preserve structural properties. 
Li et al.~\cite{li2025dynamic} modeled temporal evolution patterns using a Hidden Markov Graph (HMG) and inject noise into model parameters under differential privacy to generate synthetic graphs that retain dynamic structural characteristics.

While these approaches effectively provide privacy guarantees and preserve key structural properties, they primarily focus on the design of privacy-preserving mechanisms. 
In contrast, the vulnerability of such systems to poisoning attacks during the data publishing process remains largely underexplored, potentially introducing significant security risks under adversarial manipulation.

\begin{table*}[t]
\centering
\caption{A comprehensive comparison between existing poisoning attacks and our scheme.}
\label{tab:comparison}
\begin{tabularx}{0.70\textwidth}{llll>{\raggedright\arraybackslash}X}
\toprule
Scheme & Graph & Privacy & Attack Level & Target Task \\
\midrule
Nettack \cite{zugner2018nettack} 
& Static & -- & Edge & Node classification \\
Meta-Attack \cite{zugner2019adversarial} 
& Static & -- & Edge & Node classification \\
RL-S2V \cite{dai2018adversarial} 
& Static & -- & Edge & Node / graph classification \\
NIPA \cite{sun2020adversarial} 
& Static & -- & Node & Node classification \\
TDGIA \cite{li2020tdgia} 
& Static & -- & Node & Node classification \\
Dyn-PGD \cite{sharma2022imperceptible} 
& Dynamic & -- & Edge & Dynamic graph models \\
GDA \cite{chen2024dynamicattack}
& Dynamic & -- & Edge & Dynamic graph embedding \\
MemFreezing \cite{dai2025memfreezing} 
& Temporal & -- & Node / Edge & Temporal graph learning \\
RVA/RNA/MGA \cite{he2025data} 
& Static & LDP & User / Node & Graph statistics \\
Fake User Injection \cite{cao2021ldp} 
& Tabular & LDP & User & Frequency estimation / heavy hitters \\
OPA \cite{li2023opa} 
& Tabular & LDP & User & Mean / variance estimation \\
\textbf{Our scheme} 
& \textbf{Dynamic} & \textbf{LDP} & \textbf{Node / Edge} & \textbf{Graph statistics} \\
\bottomrule
\end{tabularx}
\end{table*}

\subsection{Data Poisoning Attacks on Dynamic Graphs}

Recent studies on graph data poisoning primarily focus on graph learning models, especially graph neural networks (GNNs), by introducing structural perturbations. 
For instance, Zügner et al.~\cite{zugner2018nettack} proposed Nettack, which misleads node classification by modifying a small number of edges or node features around a target node. Zügner and Günnemann~\cite{zugner2019adversarial} further introduced Meta-Attack, which leverages meta-learning to simulate the training process and identify globally effective perturbations. Dai et al.~\cite{dai2018adversarial} proposed RL-S2V, which employs reinforcement learning to automatically learn attack strategies, improving adaptability. In addition to modifying existing structures, Sun et al.~\cite{sun2020adversarial} proposed NIPA, which injects malicious nodes and connects them to target nodes to mislead predictions, while Li et al.~\cite{li2020tdgia} further optimize the connection strategies to enhance attack effectiveness.
With the development of temporal models, poisoning attacks have also been extended to dynamic graph learning tasks. 
For example, Sharma et al.~\cite{sharma2022imperceptible} proposed Dyn-PGD, which progressively perturbs edges across time steps, and Chen et al.~\cite{chen2024dynamicattack} introduced GDA to identify critical edges using gradient information. For temporal graph neural networks, Dai et al.~\cite{dai2025memfreezing} proposed MemFreezing, which degrades model performance by disrupting memory mechanisms.

Beyond graph learning models, recent studies have explored poisoning attacks in statistical analysis under LDP. Cao et al.~\cite{cao2021ldp} provided the first systematic investigation of poisoning attacks against LDP protocols, demonstrating that attackers can manipulate statistical results by injecting fake users. Building on this, Li et al.~\cite{li2023opa} proposed OPA attacks, enabling more fine-grained manipulation of statistical outputs. In the context of graph data, He et al.~\cite{he2025data} extended such attacks to LDP-based graph analysis protocols and designed multiple strategies (e.g., RVA, RNA, and MGA) to manipulate graph statistics such as degree centrality and clustering coefficient by forging adjacency information.

As shown in Table~\ref{tab:comparison}, these works primarily focus on either adversarial attacks against graph learning models or poisoning attacks in non-graph or static graph settings under LDP.
However, poisoning attacks against LDP-based dynamic graph statistics remain largely unexplored. 
In particular, under decentralized settings, how to characterize the impact of attacks on released data and design effective temporal attack strategies have not been systematically studied. This gap motivates the focus of this work.


\section{Preliminary}

\label{sec:preliminary}

In this section, we present the formal definitions of dynamic graphs and local differential privacy (LDP). 
Key notations used throughout the paper are summarized in Table \ref{tab:notations}.

\begin{table*}[t]
\centering
\caption{Table of Notations.}
\label{tab:notations}
\begin{tabularx}{0.7\textwidth}{l>{\raggedright\arraybackslash}X}
\toprule
Symbol & Definition \\
\midrule
$G,\; G',\; G_{HMG},\; G_{HMG}'$ 
& Original graph, attacked graph, perturbed graph, perturbed attacked graph \\

$t_i, T$ 
& Time step and the time-step set \\

$V(G_{t_i}), E(G_{t_i}), V_{t_i}', E_{t_i}^+$ 
& Real node/edge set and Fake node/edge set at $t_i$ \\

$v, v', V_{goal}$ 
& Real node, fake node, target node set \\

$m, m_{t_i}, M$ 
& Total fake-node number, number at $t_i$, number set \\

$\beta, \gamma$ 
& Fake node ratio and target node ratio \\

$\theta, \alpha$ 
& Fake node scale scheduling strategy and topology attack strategy \\

$w, w_i, W_{total}$ 
& Time decay coefficient, weight at $t_i$, total weight \\

$\overline{d}_{t_i}, D$ 
& Average degree at $t_i$ and degree set \\

$\varepsilon, \varepsilon_1, \varepsilon_2, \varepsilon_3$ 
& Total and stage-wise privacy budgets \\

$Ind_{r}^{t_i}, Ind_{f}^{t_i}$ 
& Topological indices of target nodes in real and perturbed graphs \\

$Gain(t_i), Gain$ 
& Time-step and total privacy leakage gain \\
\bottomrule
\end{tabularx}
\end{table*}

\subsection{Dynamic Graphs and Metrics}
A dynamic graph is defined as a sequence of static graph snapshots $\mathcal{G} = \{G_1, G_2, \dots, G_T\}$ evolving over a set of discrete time steps $t \in \{1, \dots, T\}$. Each snapshot $G_t = (V, E_t)$ consists of a fixed set of users (nodes) $V = \{v_1, \dots, v_N\}$ and a time-varying set of edges $E_t \subseteq V \times V$. An edge $(u, v) \in E_t$ indicates an interaction between user $u$ and user $v$ at time $t$.
In this work, we focus on four fundamental graph metrics for evaluating attacker's performance:

\textbf{Degree Centrality (DC).} 
Degree centrality measures the importance of a node based on its direct connectivity, reflecting how actively the node interacts with other nodes in the network. In dynamic graphs, $DC_t(v)$ captures the evolution of node influence over time, as the degree of a node may change over time due to the addition or removal of edges. A higher value of degree centrality indicates that the node has more direct connections and thus plays a more central role in the network at time $t$.
For a specific node $v$ at time $t$, the normalized degree centrality is defined as:
\begin{equation}
    DC_t(v) = \frac{\text{deg}_t(v)}{N - 1}
\end{equation}
where $\text{deg}_t(v)$ denotes the number of edges incident to $v$ in $G_t$, and $N$ is the total number of nodes.

\textbf{Clustering Coefficient (CC).}
The clustering coefficient measures the local cohesiveness of a node by quantifying how densely its neighbors are connected. It reflects the tendency of a node to form tightly knit groups or communities. In dynamic graphs, $CC_t(v)$ captures the temporal variation of local structural patterns, as the connections among a node’s neighbors may evolve over time due to edge additions or deletions. A higher clustering coefficient indicates that the neighbors of node $v$ are more interconnected, suggesting stronger local clustering behavior at time $t$.
For a specific node $v$ at time $t$, the local clustering coefficient is defined as:
\begin{equation}
    CC_t(v) = \frac{2 \cdot e_t(v)}{\text{deg}_t(v)\left(\text{deg}_t(v)-1\right)}
\end{equation}
where $e_t(v)$ is the number of edges between the neighbors of $v$ in $G_t$.

\textbf{Triangle Count (TC).}
Triangle count measures the number of closed triplets involving a node, capturing higher-order structural interactions beyond pairwise connections. It reflects the extent to which a node participates in tightly connected substructures. In dynamic graphs, $TC_t(v)$ characterizes the evolution of such higher-order patterns, which may change as edges are added or removed over time. A larger triangle count indicates stronger local connectivity and a higher likelihood of the node being part of cohesive groups at time $t$.
For a specific node $v$ at time $t$, the triangle count is defined as:
\begin{equation}
    TC_t(v) = 
    \left| \left\{ (u,w) \mid (v,u), (v,w), (u,w) \in E_t \right\} \right|
\end{equation}
which measures the number of triangles that include node $v$.

\textbf{Average Shortest Path Length (ASPL).}
The average shortest path length measures the average distance between a node and all other reachable nodes, reflecting its global position in the network. It captures how efficiently information can propagate from the node to others. In dynamic graphs, $ASP_t(v)$ reflects the temporal evolution of global connectivity, as changes in edges may alter shortest paths across the network. A smaller value indicates that node $v$ is more centrally located and can reach other nodes more efficiently at time $t$.
For a target node $v$, we first compute the set of reachable nodes $R_v$ using breadth-first search. 
The average shortest path length of $v$ at time $t$ is defined as:
\begin{equation}
    ASP_t(v) =
    \begin{cases}
    \frac{1}{|R_v|-1} \sum\limits_{u \in R_v,\, u \neq v} d_t(v,u), & |R_v| > 1 \\
    0, & |R_v| \le 1
    \end{cases}
\end{equation}
where $d_t(v,u)$ denotes the unweighted shortest path distance between nodes $v$ and $u$ in $G_t$.

To quantify structural changes caused by attacks, we define the ASPL-based gain for target nodes as:
\begin{equation}
    \text{shortest\_path\_gain} =
    \sum_{v \in \text{targets}}
    \left|
    ASP_{\text{attacked}}(v) -
    ASP_{\text{original}}(v)
    \right|.
\end{equation}

\subsection{LDP Protocols for Dynamic Graphs}
Local differential privacy (LDP) \cite{lin2022towards,ye2020lf,xu2025continuous} provides strong privacy guarantees by ensuring that an untrusted aggregator cannot infer sensitive information from individual user reports. 
Unlike central differential privacy, where noise is injected after data collection, LDP performs randomized perturbation directly on the user side before data is shared, thereby eliminating the need for a trusted data curator.
\begin{definition}[$\epsilon$-Local Differential Privacy]
A randomized algorithm $\mathcal{A}$ satisfies $\epsilon$-LDP if and only if for any pair of input values $x, x'$ in the domain of $\mathcal{A}$, and for any output $y \in \text{Range}(\mathcal{A})$:
\begin{equation}
    \Pr[\mathcal{A}(x) = y] \le e^\epsilon \cdot \Pr[\mathcal{A}(x') = y]
\end{equation}
where $\epsilon > 0$ is the privacy budget. A smaller $\epsilon$ indicates stronger privacy protection but lower data utility.
\end{definition}

In the context of graph data, users typically employ the randomized response (RR) mechanism to perturb binary edge information. For an edge existence bit $b \in \{0, 1\}$, the RR mechanism outputs $b'$ such that:
\begin{equation}
    \Pr[b' = 1 | b = 1] = \frac{e^\epsilon}{1 + e^\epsilon}, \quad \Pr[b' = 1 | b = 0] = \frac{1}{1 + e^\epsilon}
\end{equation}

Recent studies have proposed several LDP-based mechanisms for dynamic graph data protection \cite{dwork2006calibrating, erlingsson2014rappor, hou2023ppdu, yuan2024psgraph, hao2025caldp, li2025dynamic}. 
Among them, the Hidden Markov Graph (HMG) framework \cite{li2025dynamic} represents a state-of-the-art approach that models the temporal evolution of graphs while enforcing local differential privacy.
In HMG, each user independently perturbs their local graph information at each time slice before reporting to the server. 

The perturbation process consists of three components: (1) Degree perturbation, where users add Laplace noise to their node degrees to protect degree information; (2) Edge randomized response, where users apply the randomized response mechanism to their local adjacency lists to protect edge existence; and (3) Transition probability perturbation, where users compute the transition counts of edges between consecutive timestamps (e.g., $0\to0$, $0\to1$) and add Laplace noise to these counts.
After collecting the perturbed reports from all users, the server aggregates the noisy statistics and employs a Hidden Markov Model (HMM) to estimate the underlying dynamic graph sequence.

\section{Poisoning Attacks to LDP for Dynamic Graphs}
\label{sec:method}

\subsection{Threat Model}
To systematically analyze the potential threats posed by poisoning attacks on dynamic graph metric estimation under LDP constraints, we characterize the threat model from two dimensions: attack assumption and attack goal.

\textbf{Attack Assumption.}
We assume that the attacker can control a subset of users in each time slice under the LDP protocol, and these compromised fake users are able to submit arbitrarily crafted data to the central server. 
Formally, suppose that in time slice $G_{t_i}$ there are $n_{t_i}$ genuine users in the system, while the attacker controls $m_{t_i}$ fake users. 
Then, the total number of users participating in the protocol during this time slice is $N_{t_i} = n_{t_i} + m_{t_i}$. 
In fact, for real-world application scenarios, an attacker may acquire and manipulate a large number of fake users by purchasing accounts at low cost or compromising end devices through malware. 
Although the service provider may detect and suspend some abnormal accounts in certain time slices, the attacker can still control fake users whose numbers and identities are not entirely the same across different time slices, thereby posing a long-term and stealthy security threat to the system.

Since LDP protocols perform randomized perturbation locally on the user side, the attacker can access both the implementation details and the corresponding protocol parameters. 
In particular, the attacker knows the privacy budget $\epsilon_1$, $\epsilon_2$, and $\epsilon_3$ used for perturbing node degrees, transition counts, and node adjacency vectors, respectively. 
Moreover, the attacker may possess prior statistical knowledge of the graph structure, such as the domain size of node degrees and characteristics of the perturbed dynamic graph.

\textbf{Attacker’s goal.} The goal of the attacker is to degrade the estimation performance of graph metrics associated with specific target nodes, thereby increasing the estimation error of these metrics (e.g.,  degree centrality, clustering coefficient, shortest path, and triangle count).
Formally, we denote the set of target nodes in time slice $G_{t_i}$ as $V_{\text{goal}}$. 
To achieve this objective, the attacker dynamically controls different numbers of fake users across different time slices and carefully crafts the data reported to the central server.

\subsection{Attack Gain for Dynamic Graphs}
Based on the above threat model, we further quantify the impact of poisoning attacks by defining an attack gain metric for dynamic graphs.
We denote the set of crafted adjacency information injected by malicious nodes at each time slice as 
$Y = \{y_1, y_2, \ldots, y_m\}$, 
where $y_i$ represents the adjacency information injected by the $i$-th malicious node in the corresponding time slice.
Let $\tilde{f}_{(g,b)}$ and $\tilde{f}_{(g,a)}$ denote the estimated graph metric of a target node $g$ under the local differential privacy mechanism before and after the attack, respectively. 
The gain of the target node $g$ is defined as:
\begin{equation}
    \Delta \tilde{f}_g = \left| \tilde{f}_{(g,a)} - \tilde{f}_{(g,b)} \right|, \quad \forall g \in V_{\mathrm{goal}}
\end{equation}
where $V_{\mathrm{goal}}$ denotes the set of target nodes.
Accordingly, the aggregate gain at time slice $G_{t_i}$ is defined as the sum of the gains of all target nodes at that time slice:
\begin{equation}
\mathrm{Gain}_{t_i} = \sum_{g \in V_{\mathrm{goal}}} \Delta \tilde{f}_g,i \in [0,n-1] \cap \mathbb{Z}
\end{equation}
Considering that nodes in a dynamic graph may exhibit varying levels of influence over time, and that later time slices typically carry greater significance, we introduce a temporal decay weighting mechanism to capture this dynamic characteristic. 
Specifically, let $k$ denote the weighting coefficient of the dynamic graph. 
The importance of nodes at time slice $G_{t_i}$ is defined as $k^{\,n-1-i}$,
where $n$ denotes the total number of time slices in the dynamic graph.
Accordingly, the sum of importance weights over all time slices is given by:
\begin{equation}
k_{\mathrm{goal}} = \sum_{i=0}^{n-1} k^{\,n-1-i},i \in [0,n-1] \cap \mathbb{Z}
\end{equation}
The normalized weight associated with time slice $G_{t_i}$ is therefore defined as:
\begin{equation}
w_i = \frac{k^{\,n-1-i}}{k_{\mathrm{goal}}},i \in [0,n-1] \cap \mathbb{Z}
\end{equation}
Finally, the overall gain over the entire dynamic graph is defined as the weighted sum of the aggregate gains across all time slices:
\begin{equation}
\mathrm{Gain} = \sum_{i=0}^{n-1} w_i \, \mathrm{Gain}_{t_i}.
\end{equation}

The proposed overall gain metric $\mathrm{Gain}$ comprehensively characterizes the cumulative impact of the attack on the estimation results of graph metrics for the target nodes in the dynamic graph. A larger value of $\mathrm{Gain}$ indicates a more substantial perturbation effect on the target nodes and, consequently, a higher degree of attack effectiveness.

\begin{algorithm}[t]
    \renewcommand{\algorithmicrequire}{\textbf{Input:}}
    \renewcommand{\algorithmicensure}{\textbf{Output:}}
    \caption{Overall Framework of the Proposed Algorithm}
    \label{algorithm: masp}
    \begin{algorithmic}[1]
        \Require 
            Stream graph: $G=\{G_{t_{0}}, G_{t_{1}}, \dots, G_{t_{n-1}}\}$ (including node sets $V(G_{t_i})$ and edge sets $E(G_{t_i})$),
            Time steps: $T=\{t_{0}, t_{1}, \dots, t_{n-1}\}$,
            Privacy budget: $\epsilon = \epsilon_{1} + \epsilon_{2} + \epsilon_{3}$,
            Attack strategy: $\alpha$,
            Fake node proportion: $\beta$,
            Attack scheduling strategy: $\theta$,
            Target node proportion: $\gamma$,
            Time-slice weight decay coefficient: $w$.
            
        \Ensure 
            Target node gain: $Gain$
            
            \Statex \textbf{Step 1: Target Node Selection}
            \State $n \gets \gamma \times |NodeSum(G_{t_{n-1}})|$ \Comment{Total number of target nodes}
            \State $V_{goal} \gets \text{SelectTarget}(G_{t_{n-1}}, n)$ 
          
            \State $G_{HMG} \gets \text{Perturb}(G, \epsilon, \epsilon_1, \epsilon_2, \epsilon_3)$ 
            \Comment{HMG-perturbed}
            \State $D=\{\overline d_{t_0}, \overline d_{t_1}, \dots, \overline d_{t_{n-1}}\} \gets \text{AvgDegree}(G_{HMG})$ 
            \Comment{Upper bound of attacks}
            
            \Statex \textbf{Step 2: Dynamic Selection of Fake Nodes}
            \State $m \gets \beta \times |NodeSum(G_{t_{n-1}})|$ \Comment{Total number of fake nodes}
            \State $V' \gets \text{FakeNodeControl}(m, T, \theta, w)$ 
            \Comment{\Cref{alg:schedule_strategy}}
            
            \Statex \textbf{Step 3: Fake Edge Injection}
            \State $E^{+} \gets \text{AttackGeneration}(G, V', V_{goal}, \alpha, D, T)$
            \Comment{\Cref{alg:attack_generation}}
            \State $G_{HMG}' \gets \text{Injection}(G_{HMG}, V', E^{+})$
            \Comment{Fake edge injection}
            \State $Gain \gets \text{Compute}(G_{HMG}, G_{HMG}', V_{goal}, T, w)$ 
            \Comment{\Cref{alg:compute_gain}}
            \State \Return $Gain$
    
    \end{algorithmic}
\end{algorithm}

\subsection{Dynamic Graph Data Poisoning Attack}

\subsubsection{Overall Framework of Attack Methods}
In this paper, we propose an attack method for privacy-preserving dynamic graphs, which influences the gain of target nodes by injecting fake edges. The overall framework is illustrated in  Fig.~\ref{fig:attack_framework}, and the attack process mainly consists of three stages:

\textbf{(1) Target Node Selection:} The target node set is selected from the last time slice of the dynamic graph. All subsequent attacks are performed based on this set, and the gain of graph metrics is evaluated exclusively on these target nodes. To ensure the generality of the proposed method, attack target nodes are randomly selected.

\textbf{(2) Dynamic Selection of Fake Nodes:} At each time step, the attacker dynamically controls a number of fake nodes according to the attack strategy $\alpha$, where both the quantity and distribution are determined by the strategy. 
These controlled nodes perform connection-based attacks on the target nodes within the corresponding time slice. To ensure general applicability, fake nodes are randomly selected from the node set.

\textbf{(3) Fake Edge Injection:} The attacker uploads forged connection information through controlled fake nodes, i.e., constructs fake edges associated with target nodes to alter their graph metrics and achieve the attack objective. 

The specific procedure is described in Algorithm \ref{algorithm: masp}. First, a set of target nodes $V_{\text{goal}}$ is randomly selected from the last time slice $G_{t_{n-1}}$ according to the ratio of the target nodes $\gamma$. Then, HMG perturbation is applied to the original dynamic graph $G$, and the average degree $\overline{d}_{t_i}$ of the perturbed graph at each time slice is computed as the upper bound on the number of edges that fake nodes can establish. Based on this, the total number of fake nodes $m$ is determined according to the ratio of fake nodes $\beta$. Combined with the time steps $T$, the attack scheduling strategy $\theta$ and the temporal decay factor $w$, the number of fake nodes in each time slice is dynamically controlled. Subsequently, under the joint influence of the attack strategy $\alpha$, the target node set $V_{\text{goal}}$, and both structural and temporal information of the graph, a fake edge set $E^{+}$ is generated and injected into the perturbed dynamic graph, resulting in the attacked graph $G'_{\text{HMG}}$. In practice, fake nodes $v'$ upload forged adjacency vectors to stabilize and enhance the attack effect, while real nodes $v$ still upload their perturbed true adjacency vectors, thereby achieving effective attacks under privacy constraints. Finally, the overall gain of the target nodes is computed by incorporating both the target node set and temporal decay weights. Overall, this method forms a complete attack pipeline, including target node selection, fake node scheduling, fake edge injection, and gain evaluation.

Dynamic graphs released by systems are typically subjected to noise injection mechanisms to ensure privacy protection. Unlike static graphs, noise injection in dynamic graphs must not only perturb structural information but also account for the temporal correlations introduced by the evolution of nodes and edges over time. In this work, we adopt the HMG method, which is specifically designed for dynamic graph perturbation, to process the data, thereby enhancing privacy protection while preserving the utility of graph structural information. It is worth noting that applying noise to the adversarial data uploaded by fake users would significantly weaken the attack effectiveness; therefore, this portion of the data is not perturbed in our approach.

\subsubsection{Dynamic Selection of Fake Nodes}

The number of fake nodes has a significant impact on the estimation of graph metrics under the same attack strategy, as it directly determines the attack intensity and its propagation range within the graph, thereby influencing the magnitude of estimation bias. Therefore, properly determining the number of fake nodes is crucial for achieving effective attacks. To this end, Algorithm \ref{alg:schedule_strategy} theoretically constrains the range of fake node quantities under different strategies and characterizes the dynamic scheduling mechanism of fake node scales across time slices, thereby providing a unified and controllable basis for comparing the attack performance of different strategies.

\begin{algorithm}[t]
\caption{FakeNodeControl: Dynamic Selection of Fake Nodes}
\label{alg:schedule_strategy}
\begin{algorithmic}[1]
\Require Time steps: $T=\{t_0,t_1,\dots,t_{n-1}\}$, Total number of fake nodes: $m$, Attack scheduling strategy: $\theta \in \{0,1,2\}$, Time-slice weight decay coefficient: $w$
\Ensure Controlled fake-node sets: $V'$

\If{$\theta = 0$} \Comment{Average attack scheduling strategy}
    \State $m_{t_i} \gets m/n,\ \forall i$

\ElsIf{$\theta = 1$} \Comment{Random attack scheduling strategy}
    \State $\mathcal{R} \gets \text{Sample}(\{1,\dots,m-1\},n-1)$ \Comment{Random sampling without replacement}
    \State $(s_1,\dots,s_{n-1}) \gets \text{Sort}(\mathcal{R})$
    \State $s_0 \gets 0,\ s_n \gets m$ \Comment{Set boundaries}
    \State $m_{t_i} \gets s_{i+1}-s_i,\ \forall i$

\ElsIf{$\theta = 2$} \Comment{Decay-function-based attack scheduling strategy}
    \State $w_i \gets w^{\,n-1-i},\ \forall i$ \Comment{Calculate time-slice weight}
    \State $W_{total} \gets \sum_{i=0}^{n-1} w_i$ \Comment{Calculate total weight}
    \State $m_{t_i} \gets \lfloor (w_i/W_{total}) \times m \rfloor,\ \forall i$ \Comment{Number of fake nodes per time slice}
\EndIf

\State $V'_{t_i} \gets \text{NodeControl}(G_{t_i}, V_{goal}, m_{t_i}),\ \forall i$
\State $V' \gets\{V'_{t_0},V'_{t_1},\dots,V'_{t_{n-1}}\},\ \forall i$

\State \Return $V'$
\end{algorithmic}
\end{algorithm}

The algorithm first determines the number of fake nodes in each time slice $m_{t_i}$ according to the selected strategy. When $\theta=0$, average attack scheduling strategy is adopted, where each time slice is assigned the same number of fake nodes; when $\theta=1$, random attack scheduling strategy is employed, where split points are generated via sampling without replacement to randomly divide the number of nodes across time slices; when $\theta=2$, decay-function-based attack scheduling strategy is adopted, where the weight of each time slice is computed based on $w$, and the number of fake nodes in each time slice is proportionally controlled according to these weights.

Subsequently, for each time slice, based on the controlled number of fake nodes $m_{t_i}$ and the target node set $V_{goal}$, a node control function is applied to select the corresponding fake node set $V'_{t_i}$. Finally, the fake node sets from all time slices are aggregated to obtain $V'$. This algorithm enables flexible scheduling of fake nodes under different strategies and provides a foundation for subsequent attack procedures.

\subsubsection{Fake Edge Injection}

During the process of attack edge injection, different attack strategies can significantly influence the construction of fake edges. To this end, we consider three representative attack strategies: Random Vertex Attack (RVA), Random Node Attack (RNA), and Maximum Gain Attack (MGA). Specifically, RVA establishes connections randomly within the graph to disperse the attack impact, RNA directly connects to target nodes to enhance attack specificity, and MGA adopts a structured approach to maximize the impact on target node metrics. These strategies characterize the trade-off between attack strength and targeting from different perspectives. Algorithm \ref{alg:attack_generation} describes the process of generating fake edges under different attack strategies.

\begin{algorithm}[t]
\caption{AttackGeneration: Generate Attack Edges}
\label{alg:attack_generation}
\begin{algorithmic}[1]
\Require Complete flow graph: $G=\{G_{t_0},\dots,G_{t_{n-1}}\}$, Controlled fake-node sets: $V'=\{V'_{t_0},V'_{t_1},\dots,V'_{t_{n-1}}\}$, Target nodes: $V_{goal}$, Attack strategy: $\alpha$, Set of average degrees: $D=\{\bar d_{t_0},\dots,\bar d_{t_{n-1}}\}$, Set of time steps: $T=\{t_0,\dots,t_{n-1}\}$
\Ensure the set of attacked graph: $G_{HMG}'$

\ForAll{$t_i \in T$}
    \If{$\alpha = 0$} \Comment{RVA Strategy}
    \ForAll{$V' \in V'_{t_i}$}
        \State $k \gets \text{RandInt}(1,\bar d_{t_i})$ \Comment{Number of injected edges in $[1,\bar d_{t_i}]$}
        \State $ E^{+}_{t_i} \gets E^{+}_{t_i} \cup \{(v',v)\} $ , where $v \in \text{Sample}(V(G_{t_i}),k)$
    \EndFor

    \ElsIf{$\alpha = 1$} \Comment{RNA Strategy}
        \ForAll{$v' \in V'_{t_i}$}
            \State $E^{+}_{t_i} \gets E^{+}_{t_i} \cup \{(v',v)\}$, where $v \in \text{RandSelect}(V_{goal},1)$  \Comment{Number of injected edges = 1}
        \EndFor
    
    \ElsIf{$\alpha = 2$} \Comment{MGA Strategy}
        \State $Pairs \gets \text{PairFakeNode}(V'_{t_i},\bar d_{t_i})$ \Comment{Stage 1: Pair fake nodes}
        \ForAll{$(u',v') \in Pairs$}  \Comment{Stage 2: Connect to common target nodes}
            \State $E^{+}_{t_i} \gets E^{+}_{t_i} \cup \text{Connect}(u',v',V_{goal},\bar d_{t_i})$
        \EndFor
    \EndIf
\EndFor

\State $E^{+} \gets \{E^{+}_{t_0}, E^{+}_{t_1}, \dots, E^{+}_{t_{n-1}}\}$,$\ \forall i$

\State $G_{HMG}' \gets \text{Injection}(G_{HMG}, V', E^{+})$
\Comment{Fake edge injection}

\State \Return $G_{HMG}'$
\end{algorithmic}
\end{algorithm}

\begin{algorithm}[t]
    \renewcommand{\algorithmicrequire}{\textbf{Input:}}
    \renewcommand{\algorithmicensure}{\textbf{Output:}}
    \caption{Compute: Attack Gain Computation}
    \label{alg:compute_gain}
    \begin{algorithmic}[1]
        \Require 
            Complete ground-truth flow graph sequence: $G_{HMG} = \{G_{HMG_{t_0}}, G_{HMG_{t_1}}, \dots, G_{HMG_{t_{n-1}}}\}$,
            Set of time steps: $T = \{t_0, \dots, t_{n-1}\}$,
            Decay weight: $w$,
            Set of target nodes: $V_{goal}$,
            Attacked graph sequence: $G'_{HMG} = \{G'_{HMG_{t_0}}, G'_{HMG_{t_1}}, \dots, G'_{HMG_{t_{n-1}}}\}$
        \Ensure 
            Total gain of graph metrics: $Gain$

        \State $Gain \gets 0$
    
        \ForAll{$t_i \in T$}
            \State $Gain(t_i) \gets 0$
            \ForAll{$v \in V_{goal}$}
                \State $Ind_r^{t_i}(v) \gets \text{Ind}(G_{HMG}[t_i], v)$  \Comment{Pre-attack graph metric value of target nodes}
                \State $Ind_f^{t_i}(v) \gets \text{Ind}(G'_{HMG}[t_i], v)$ \Comment{Post-attack graph metric value of target nodes}
                \State $Gain(t_i) \gets Gain(t_i) + \left| Ind_f^{t_i}(v) - Ind_r^{t_i}(v) \right|$ \Comment{Absolute difference in a single time slice}
            \EndFor
            \State $Gain \gets Gain + w^{n-1-i} \cdot Gain(t_i)$  \Comment{Total weighted gain}
        \EndFor
    
        \State \Return $Gain$
    \end{algorithmic}
\end{algorithm}

The algorithm controls the generation of fake edges under different attack strategies and constructs the corresponding fake edge set $E^+_{t_i}$ at each time slice $t_i$ according to the strategy $\alpha$. When $\alpha = 0$, the Random Vertex Attack (RVA) strategy is adopted, where each fake node randomly selects a number of connections between $1$ and $\overline{d}_{t_i}$, and then randomly connects to the same number of nodes from the node set of the current time slice. When $\alpha = 1$, the Random Node Attack (RNA) strategy is employed, where each fake node randomly selects one target node to establish a connection. When $\alpha = 2$, the Maximum Gain Attack (MGA) strategy is used: fake nodes are first paired, and each pair of fake nodes is then jointly connected to nodes in the target node set to influence the target nodes. Finally, the fake edge sets generated across all time slices are aggregated to obtain the overall attack edge set $E^+$, which is then injected into the original perturbed graph to produce the attacked graph $G'_{\text{HMG}}$.

In subsequent experiments, we observe that MGA outperforms both RNA and RVA in terms of attack effectiveness and stability. Therefore, to more effectively evaluate the impact of varying numbers of fake nodes across different time slices on graph metrics in privacy-preserving dynamic graphs, and to facilitate comparative analysis, the MGA strategy is primarily adopted during the attack edge generation phase in this work.
Specifically, this strategy precisely regulates the connection behavior of fake nodes, enabling them to establish as many connections as possible with target nodes and other fake nodes while reducing the risk of detection, thereby maximizing the perturbation of graph metric estimation for target nodes under given constraints.

\subsubsection{Attack scheme}

We propose three attack strategies: Equal-Scale Attack (ESA), Random-Scale Attack (RSA), and Temporal Enhanced Attack (TEA). These strategies manipulate the estimation of graph metrics for target nodes across different time slices by controlling varying numbers of fake nodes (see Algorithm \ref{alg:schedule_strategy}), which generate forged edges injected into the original adjacency matrix (see Algorithm \ref{alg:attack_generation}) and are subsequently uploaded to the central server. 

Specifically, ESA assigns an equal number of fake nodes to each time slice; RSA controls a randomly varying number of fake nodes within a predefined range across different time slices; and TEA adaptively adjusts the number of controlled fake nodes according to the importance of each time slice, thereby enabling more targeted and intensified attacks on critical time slices.

\noindent
\textbf{Equal-Scale Attack (ESA):} 
This strategy controls the same number of fake nodes to launch attacks across all time slices, thereby imposing approximately uniform attack intensity over time. ESA does not consider the relative importance of nodes across different time slices; thus, it exhibits relatively stable overall gain performance and can serve as a baseline for evaluating and comparing other attack strategies.

\noindent
\textbf{Random-Scale Attack (RSA):} 
This strategy enhances attack stealthiness by randomly controlling a certain number of fake nodes to launch attacks across different time slices. Even if the attack is detected, it is difficult to accurately infer the distribution of attack intensity over time. However, the effectiveness of RSA exhibits considerable variability. When attacks are primarily concentrated in earlier time slices, their contribution to overall metrics is relatively limited, resulting in weaker attack performance. In contrast, when attacks are concentrated in later time slices, the attack effectiveness can be significantly improved, but at the cost of an increased risk of detection.

\noindent
\textbf{Temporal-Enhanced Attack (TEA):} 
This strategy takes into account that the relative importance of nodes varies across different time slices in dynamic graphs, i.e., the importance of time slices evolves over time. As time progresses, it gradually increases the number of controlled fake nodes, thereby maximizing the impact on the estimation of graph metrics for target nodes without significantly increasing the risk of detection. Compared with ESA and RSA, TEA more effectively captures temporal evolution characteristics and achieves a better balance between attack effectiveness and stealthiness.

\subsubsection{Attack Gain Computation}

After the attack is performed, it is necessary to quantitatively evaluate its effectiveness to measure the actual impact of different strategies on target nodes in privacy-preserving dynamic graphs. To this end, we design a gain computation method for target nodes in dynamic graphs, which is based on the changes in graph metrics before and after the attack, and further incorporates temporal decay factors to comprehensively capture the influence of different time slices. Algorithm \cref{alg:compute_gain} presents the detailed computation procedure.

At each time slice $t_i$, the metric values of all target nodes are computed on both the original and attacked graphs, and the gain at time slice $t_i$, denoted as $Gain_{t_i}$, is obtained by summing the absolute differences between these values. Based on this, a temporal decay weight $w^{n-1-i}$ is introduced to perform a weighted aggregation of gains across all time slices, yielding the overall attack gain $Gain$. The final $Gain$ comprehensively reflects the effectiveness of the attack across different time slices as well as its temporal impact.

\begin{table}[b]
  \centering
  \begin{threeparttable}
  \caption{Default parameter settings.}
  \label{tab:params}
  \begin{tabularx}{\linewidth}{lc>{\raggedright\arraybackslash}X}
    \toprule 
    Parameter & Default setting & Description \\
    \midrule 
    $\beta$ & 0.05 & The fraction of fake users \\
    $\gamma$ & 0.05 & The fraction of target users \\
    $\varepsilon$ & 4.0 & Privacy budget \\
    \bottomrule 
  \end{tabularx}
  \end{threeparttable}
\end{table}

This gain computation algorithm characterizes the temporal effects of attacks within a unified framework. It not only incorporates the changes in node metrics before and after the attack on target nodes, but also introduces temporal decay weights to reflect the varying importance of different time slices, thereby better aligning with the evolutionary nature of dynamic graphs. Meanwhile, by aggregating the gains across all time slices, it enables an effective quantitative evaluation of the overall attack performance. In subsequent experiments, this gain is adopted as the evaluation metric, where a larger value indicates a more significant attack effect.

\begin{table*}[t]
  \centering
  \caption{Summary of Datasets and Graph Properties.}
  \label{tab:datasets}

  \begin{tabularx}{\textwidth}{l r r c c >{\raggedright\arraybackslash}X}
    \toprule
    \textbf{Dataset} & 
    \multicolumn{1}{c}{\textbf{Vertices}} & 
    \multicolumn{1}{c}{\textbf{Edges}} & 
    \textbf{Density} & 
    \textbf{Slices} & 
    \textbf{Description} \\
    \midrule
    \textbf{AS-733} & 3,171 & 108,357 & 0.02156 & 20 & Autonomous Systems routing graphs tracking dynamic node/edge evolution from the Univ. of Oregon Route Views. \\
    
    \textbf{Email} & 986 & 332,334 & 0.68437 & 12 & Anonymized internal email communications from a large European research institution \cite{li2025dynamic}. \\
    
    \textbf{Reply} & 27,927 & 1,096,440 & 0.00281 & 20 & Directed reply relationship networks among individuals on the Linux kernel mailing list \cite{li2025dynamic}. \\
    
    \textbf{Synthetic} & 8,484 & 176,805 & 0.00491 & 20 & Dynamic graphs generated via a modified Barabási-Albert (BA) model \cite{song2018differentially}. \\
    \bottomrule
  \end{tabularx}
\end{table*}

\begin{figure*}[t]
\setlength{\tabcolsep}{1.5pt} 
\resizebox{\textwidth}{!}{%
\begin{tabular}{@{} >{\centering\arraybackslash}m{0.3cm} @{\hspace{1pt}}
                >{\centering\arraybackslash}m{0.245\textwidth}
                >{\centering\arraybackslash}m{0.245\textwidth}
                >{\centering\arraybackslash}m{0.245\textwidth}
                >{\centering\arraybackslash}m{0.245\textwidth} @{}}

  & {\normalfont\scriptsize Ass-733}
  & {\normalfont\scriptsize Email}
  & {\normalfont\scriptsize Reply}
  & {\normalfont\scriptsize Synthetic} \\[1pt]

  \rotatebox[origin=c]{90}{\normalfont\scriptsize Gain (DC)}
  & \includegraphics[width=\linewidth]{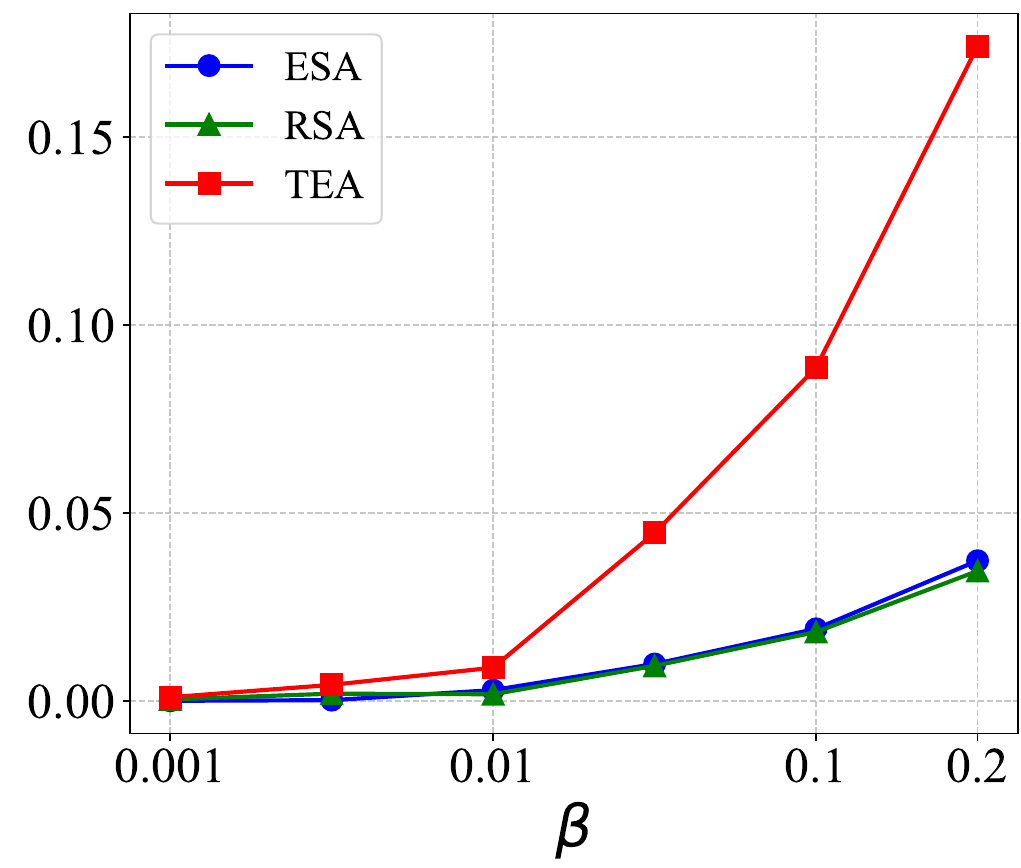}
  & \includegraphics[width=\linewidth]{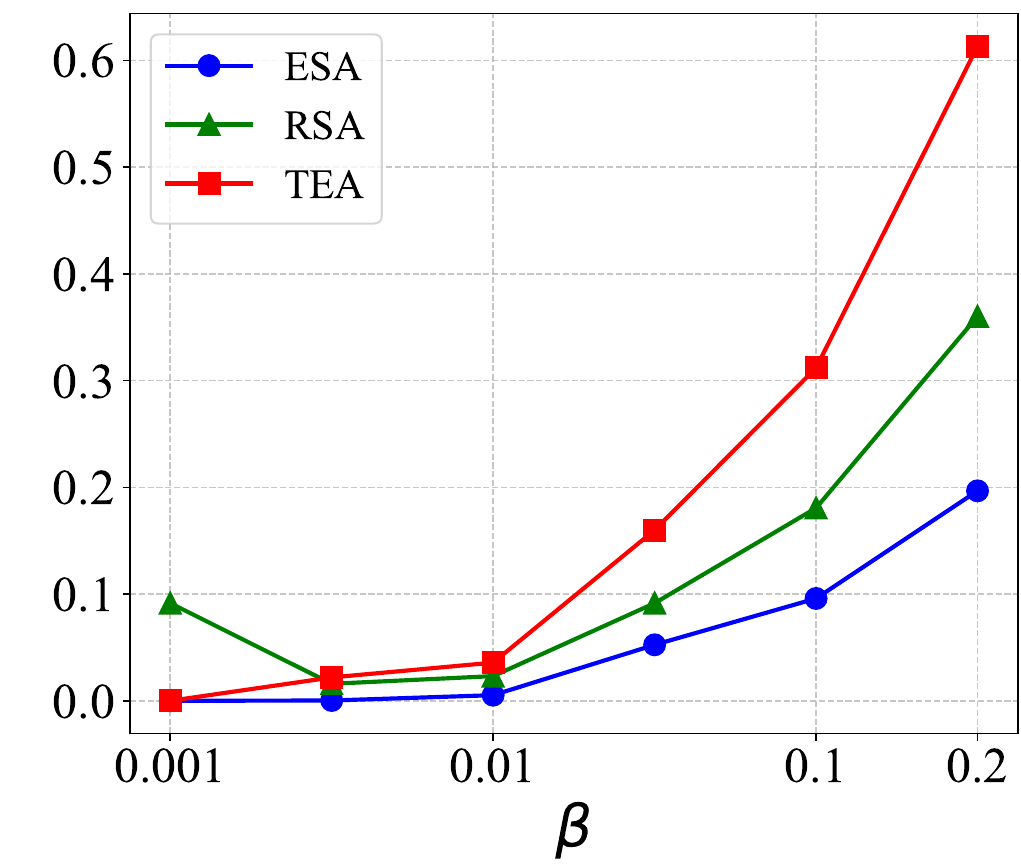}
  & \includegraphics[width=\linewidth]{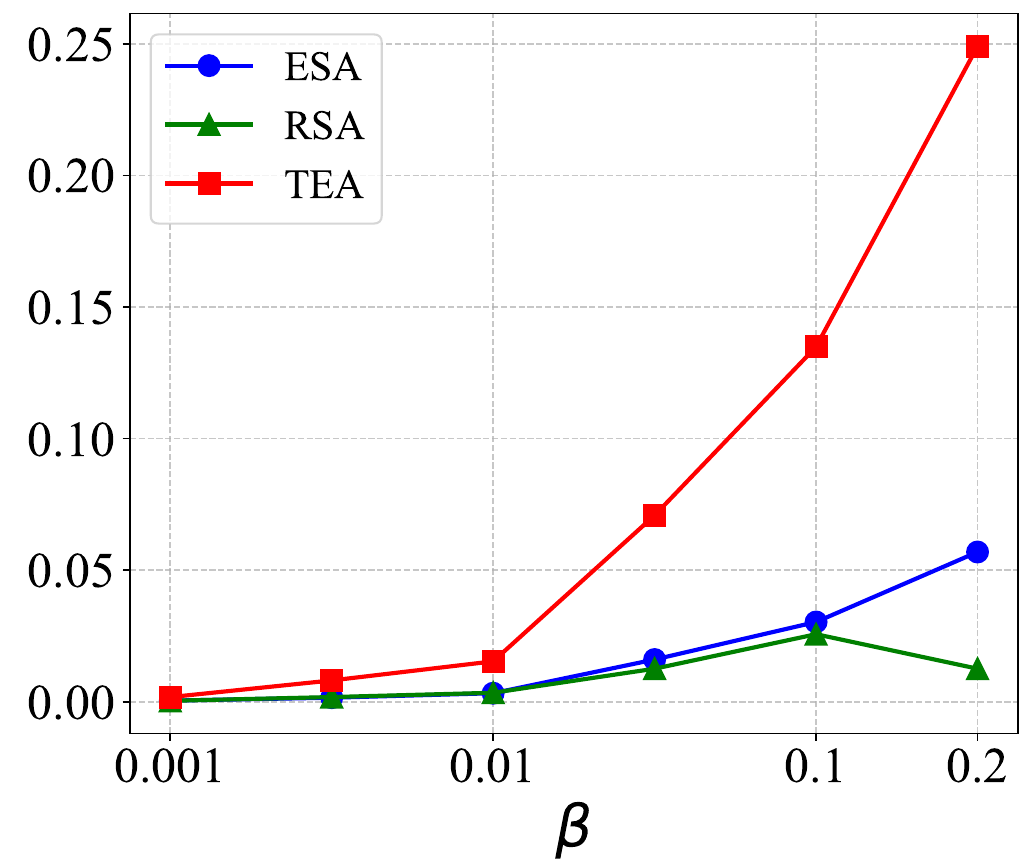}
  & \includegraphics[width=\linewidth]{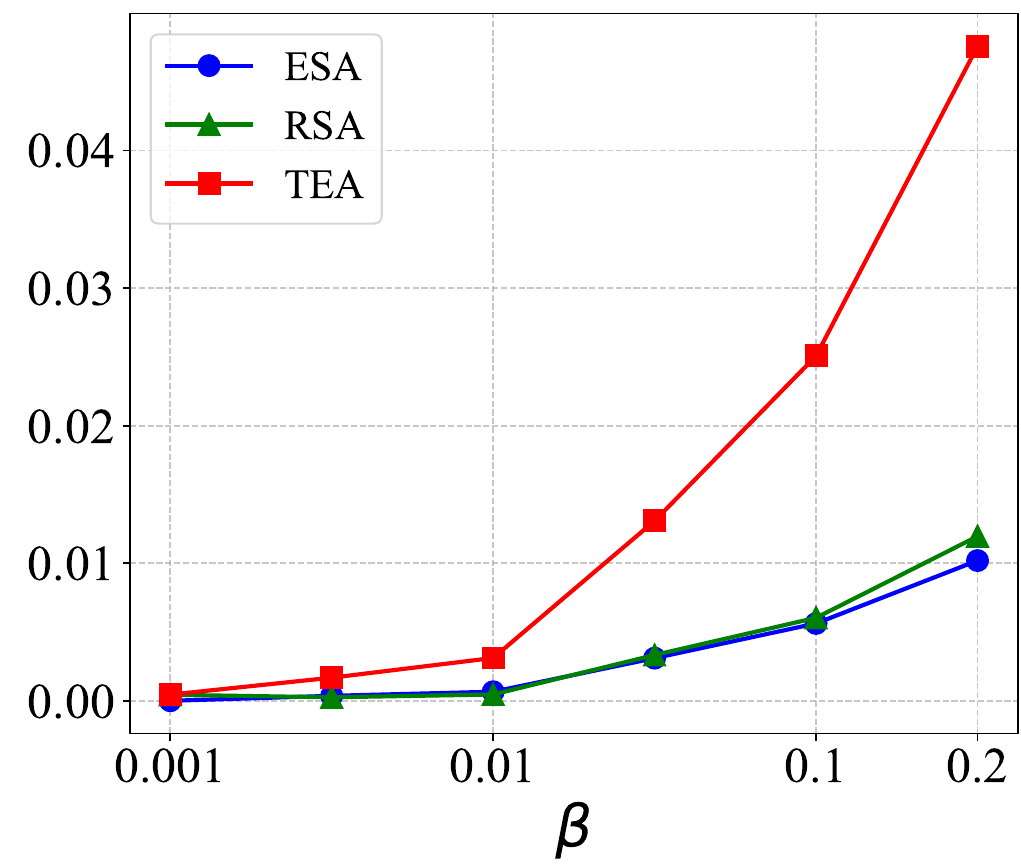} \\[-2pt]

  \rotatebox[origin=c]{90}{\normalfont\scriptsize Gain (DC)}
  & \includegraphics[width=\linewidth]{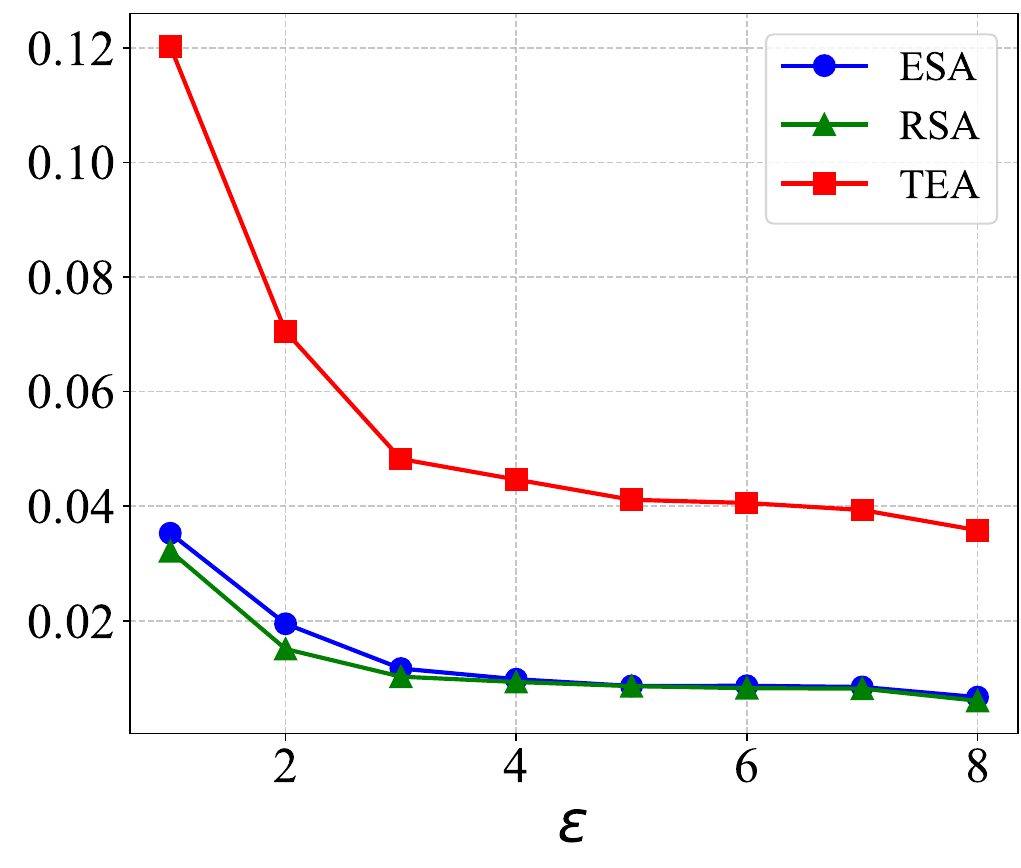}
  & \includegraphics[width=\linewidth]{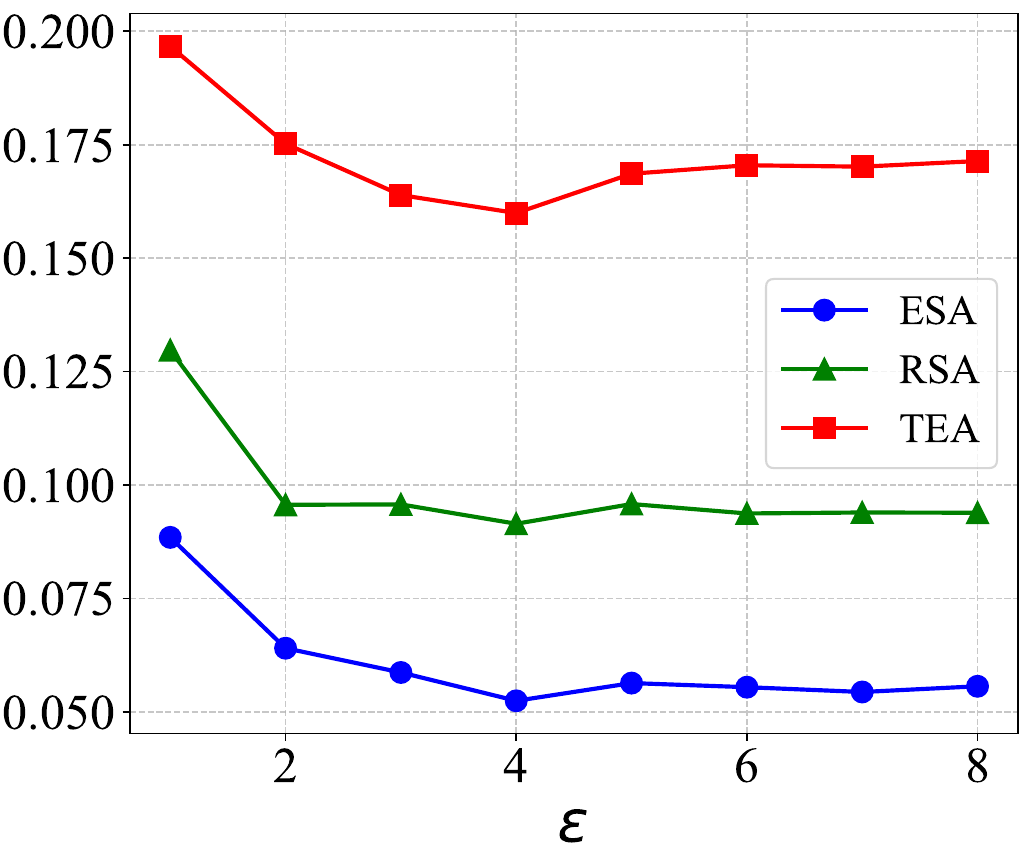}
  & \includegraphics[width=\linewidth]{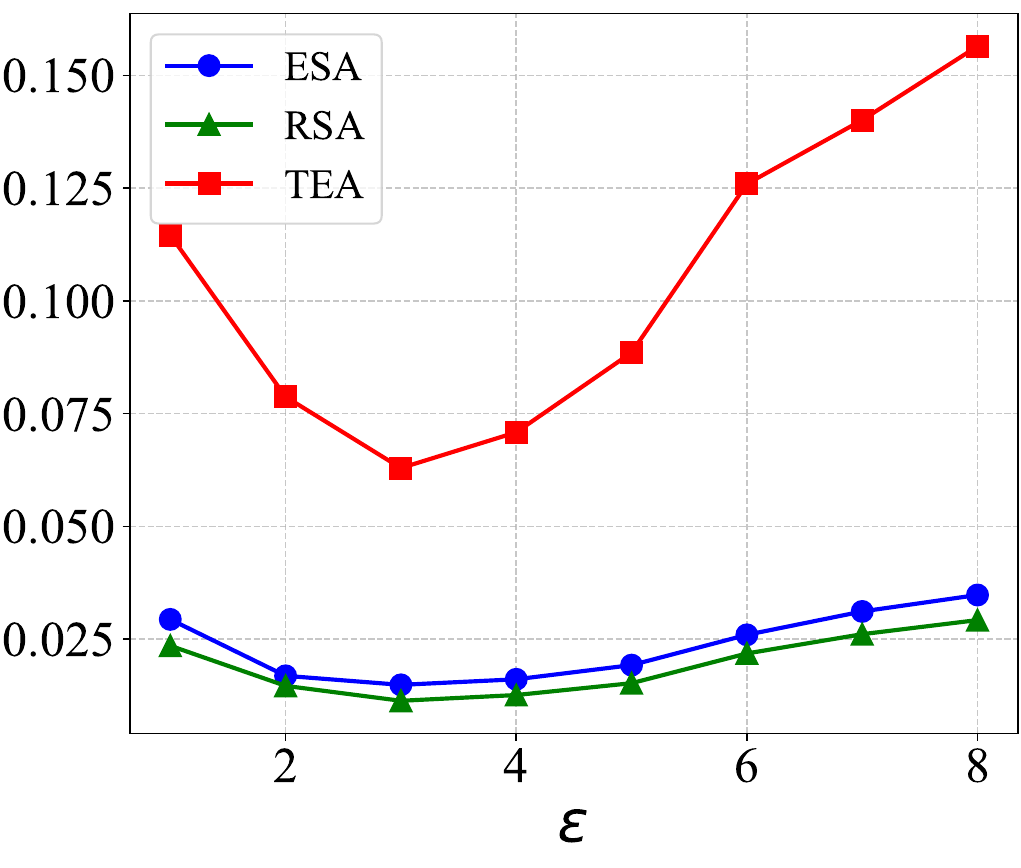}
  & \includegraphics[width=\linewidth]{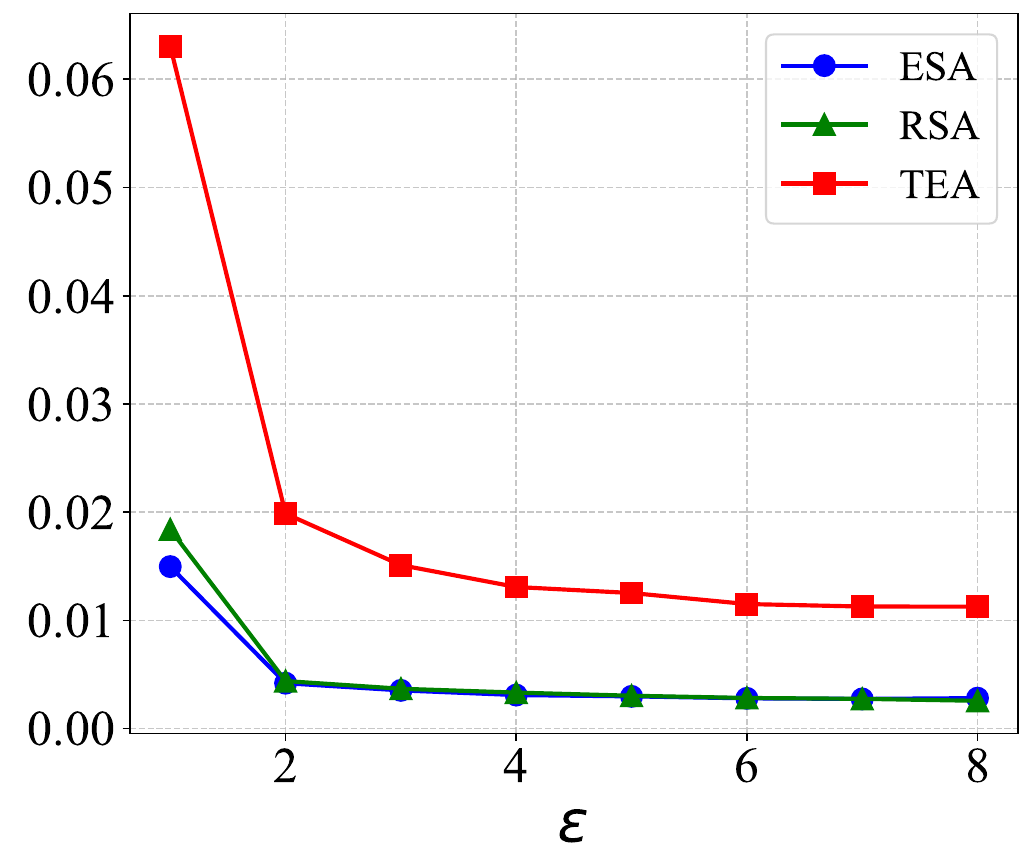} \\[-2pt]

  \rotatebox[origin=c]{90}{\normalfont\scriptsize Gain (DC)}
  & \includegraphics[width=\linewidth]{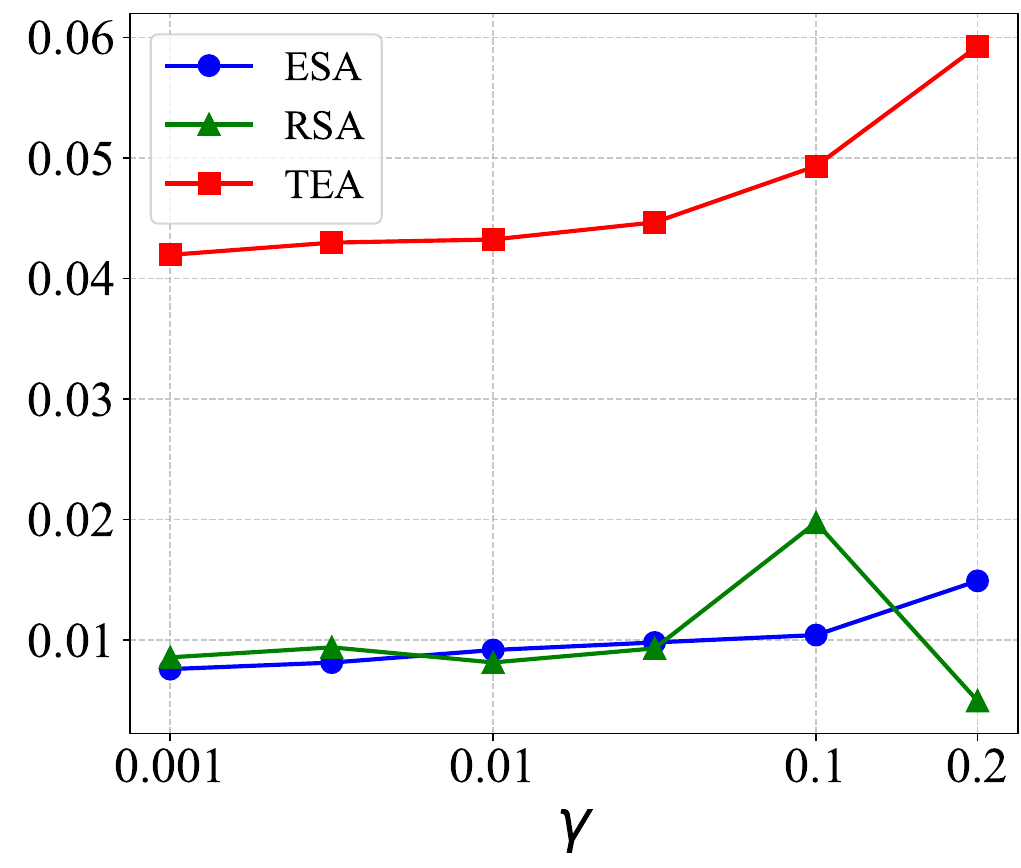}
  & \includegraphics[width=\linewidth]{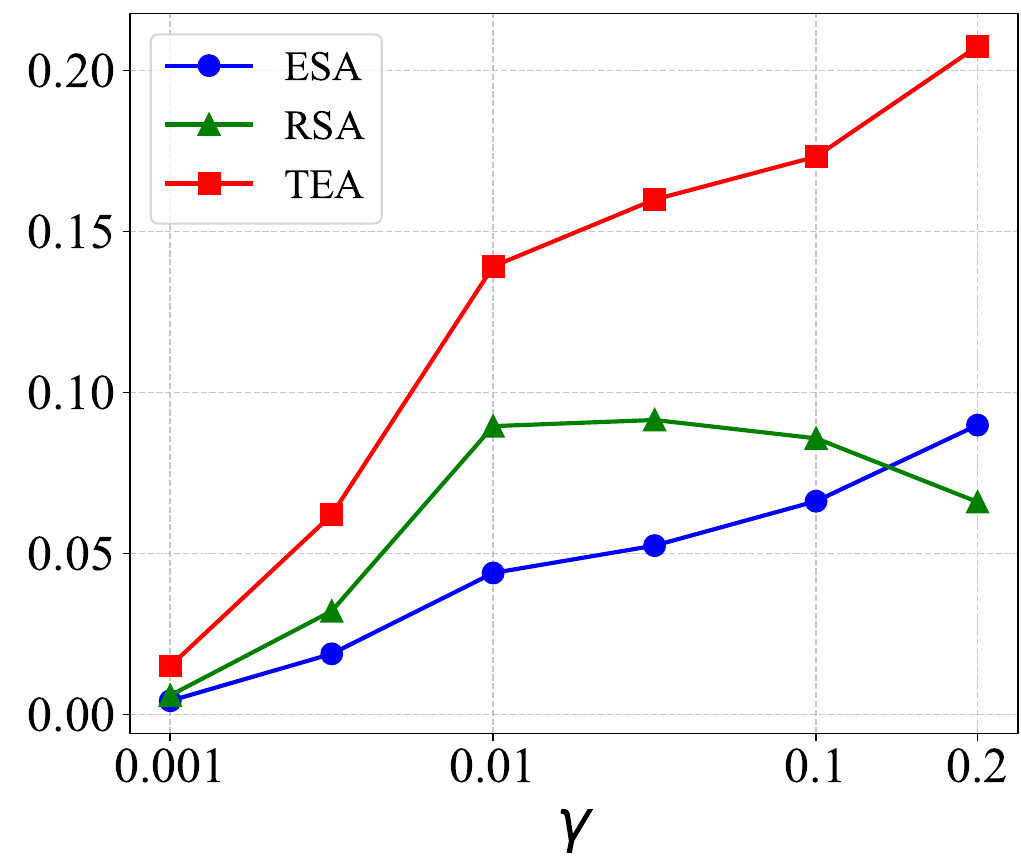}
  & \includegraphics[width=\linewidth]{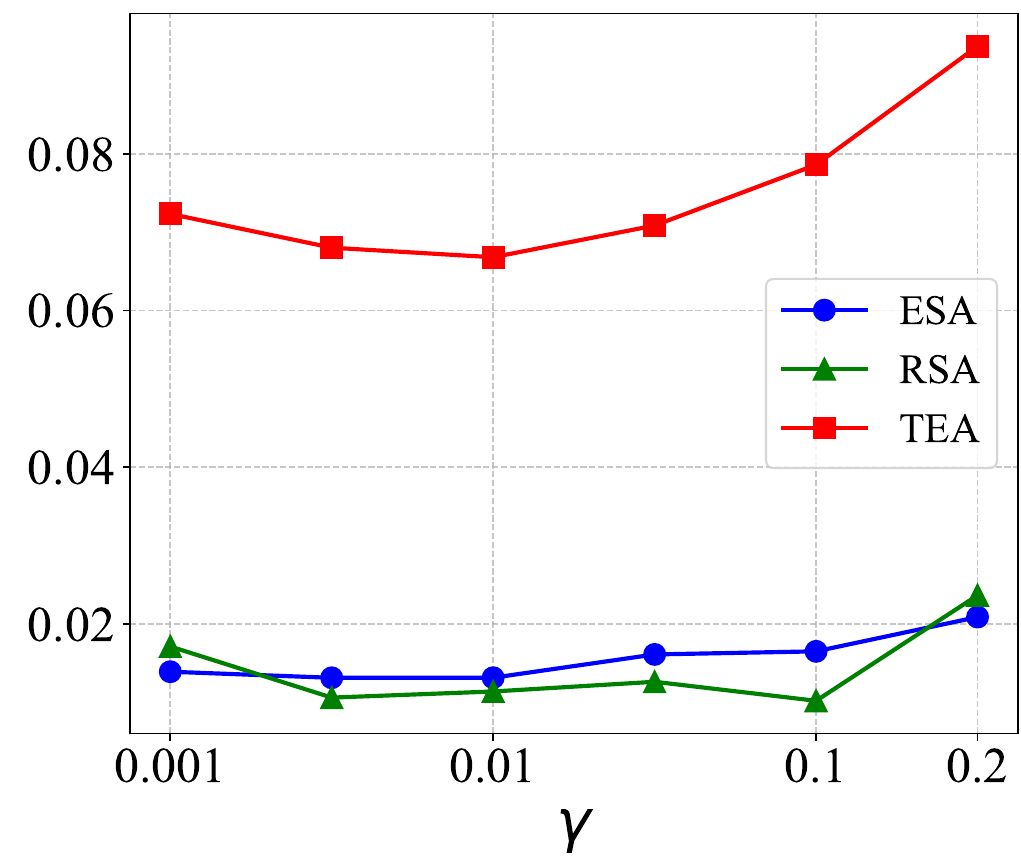}
  & \includegraphics[width=\linewidth]{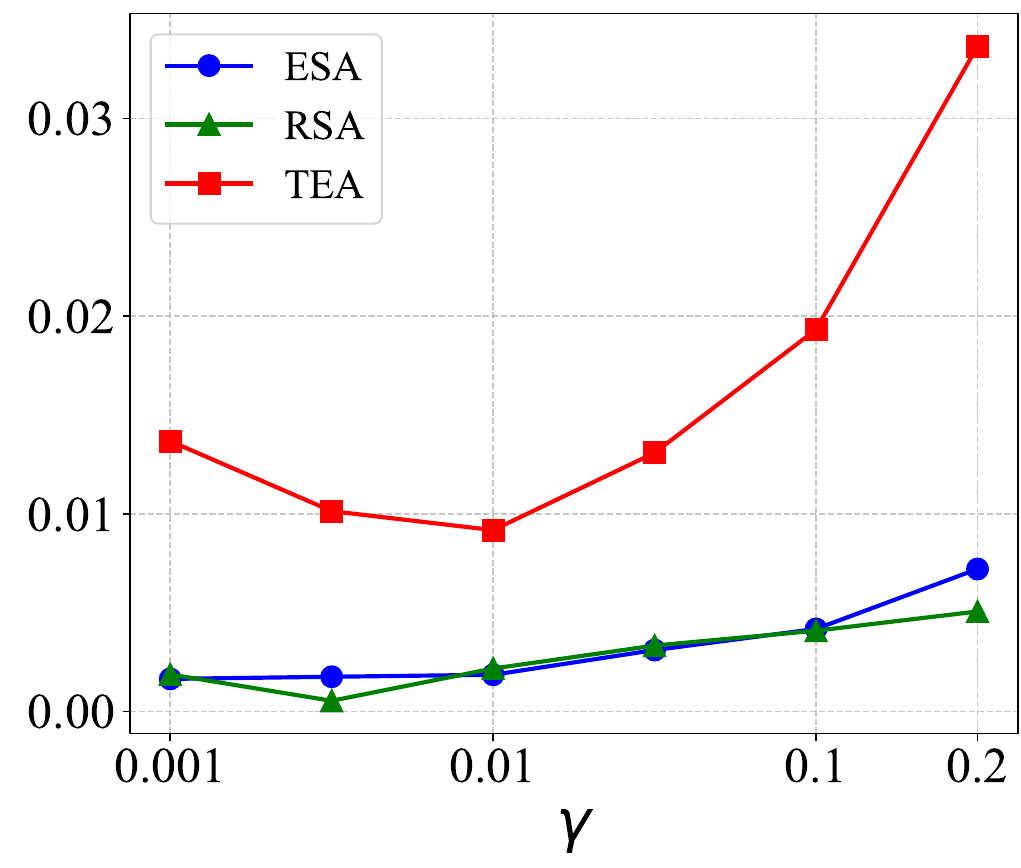} \\

\end{tabular}%
}
\caption{The effect of $\beta$, $\epsilon$ and $\gamma$ changes on gain of degree centrality}
\label{fig:degreecentrality_mga}
\end{figure*}

\section{Evaluation}
\label{sec:evaluation}
In this section, we evaluate the effectiveness of the proposed poisoning attacks on dynamic graphs. Specifically, we assess their impact on four representative graph metrics across three real-world graphs and one synthetic graph.

\subsection{Experimental Setup}

We evaluate the proposed attack on four dynamic graph datasets, comprising one synthetic and three real-world networks (AS-733, Email, and Reply). Table \ref{tab:datasets} summarizes the topological properties, number of time slices ($T$), and background details of each dataset.

The attack gain is influenced by factors including the fake node ratio $\beta$, the privacy budget $\epsilon$, and the target node ratio $\gamma$. Table~\ref{tab:params} summarizes these parameters along with their default values. In the experiments, a controlled variable approach is adopted: when evaluating the effect of a specific parameter, the remaining parameters are fixed at their default settings, enabling a systematic assessment of each parameter’s impact on the experimental results.
Specifically, $\beta \in \{0.001,\allowbreak 0.005,\allowbreak 0.01,\allowbreak 0.05,\allowbreak 0.1,\allowbreak 0.2\}$,
$\epsilon \in \{1.0,\allowbreak 2.0,\allowbreak 3.0,\allowbreak 4.0,\allowbreak 5.0,\allowbreak 6.0,\allowbreak 7.0,\allowbreak 8.0\}$,
and $\gamma \in \{0.001,\allowbreak 0.005,\allowbreak 0.01,\allowbreak 0.05,\allowbreak 0.1,\allowbreak 0.2\}$.

\subsection{Results of Degree Centrality}

Figure \ref{fig:degreecentrality_mga} illustrates the impact of varying parameters $\beta$, $\epsilon$, and $\gamma$ on the degree centrality gain of target nodes under three attack methods across four datasets. Overall, all three attacks, ESA, RSA, and TEA, are able to improve the degree centrality of target nodes, while exhibiting notable performance differences. Among them, TEA consistently achieves the best performance, whereas ESA and RSA provide relatively limited and comparable improvements. 
The effects of the three parameters are analyzed as follows:

\textbf{Fake Node Ratio $\beta$:} The first row illustrates the impact of parameter $\beta$ on the degree centrality gain. Across all four datasets, the gains of the three methods increase consistently with the growth of $\beta$, indicating that a higher proportion of fake nodes leads to more effective attacks on the degree centrality of target nodes. Among them, TEA exhibits the most significant growth and consistently outperforms ESA and RSA, while ESA and RSA show relatively moderate and similar trends. This demonstrates that $\beta$ has a positive effect on enhancing degree centrality, and TEA is more sensitive to this parameter.

\begin{figure*}[t]
\setlength{\tabcolsep}{1.5pt} 
\resizebox{\textwidth}{!}{%
\begin{tabular}{@{} >{\centering\arraybackslash}m{0.3cm} @{\hspace{1pt}}
                >{\centering\arraybackslash}m{0.245\textwidth}
                >{\centering\arraybackslash}m{0.245\textwidth}
                >{\centering\arraybackslash}m{0.245\textwidth}
                >{\centering\arraybackslash}m{0.245\textwidth} @{}}

  & {\normalfont\scriptsize Ass-733}
  & {\normalfont\scriptsize Email}
  & {\normalfont\scriptsize Reply}
  & {\normalfont\scriptsize Synthetic} \\[1pt]

  \rotatebox[origin=c]{90}{\normalfont\scriptsize Gain (TC)}
  & \includegraphics[width=\linewidth]{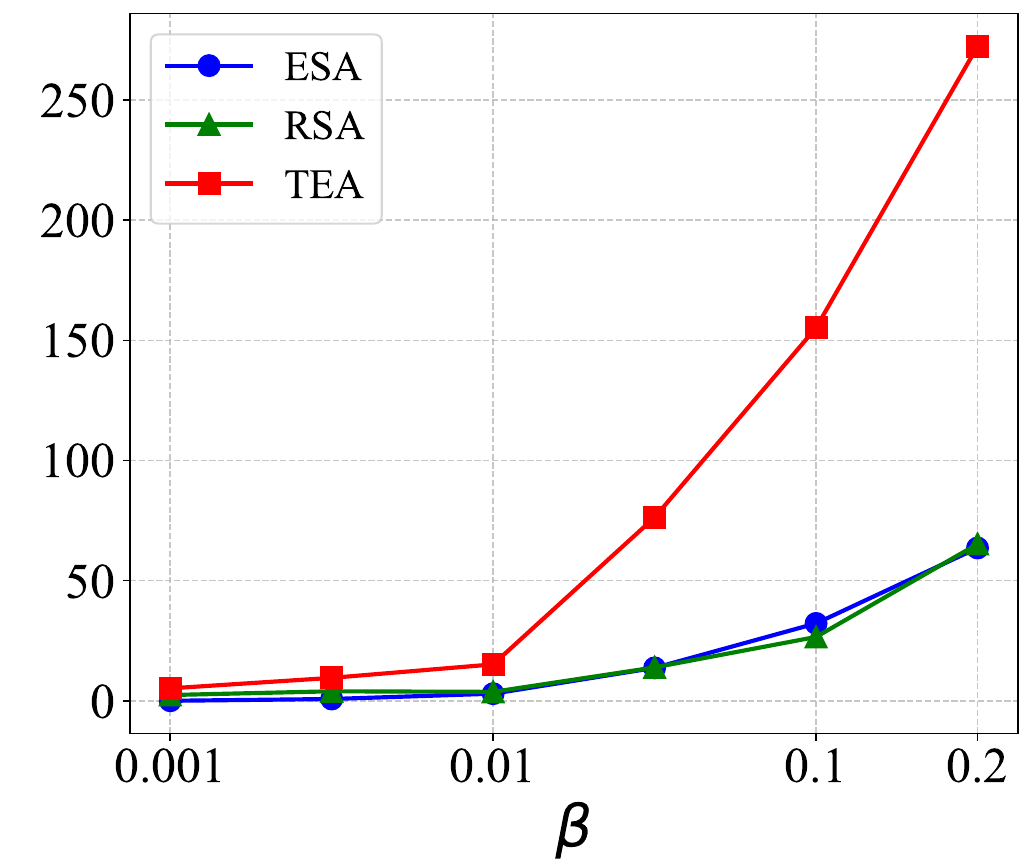}
  & \includegraphics[width=\linewidth]{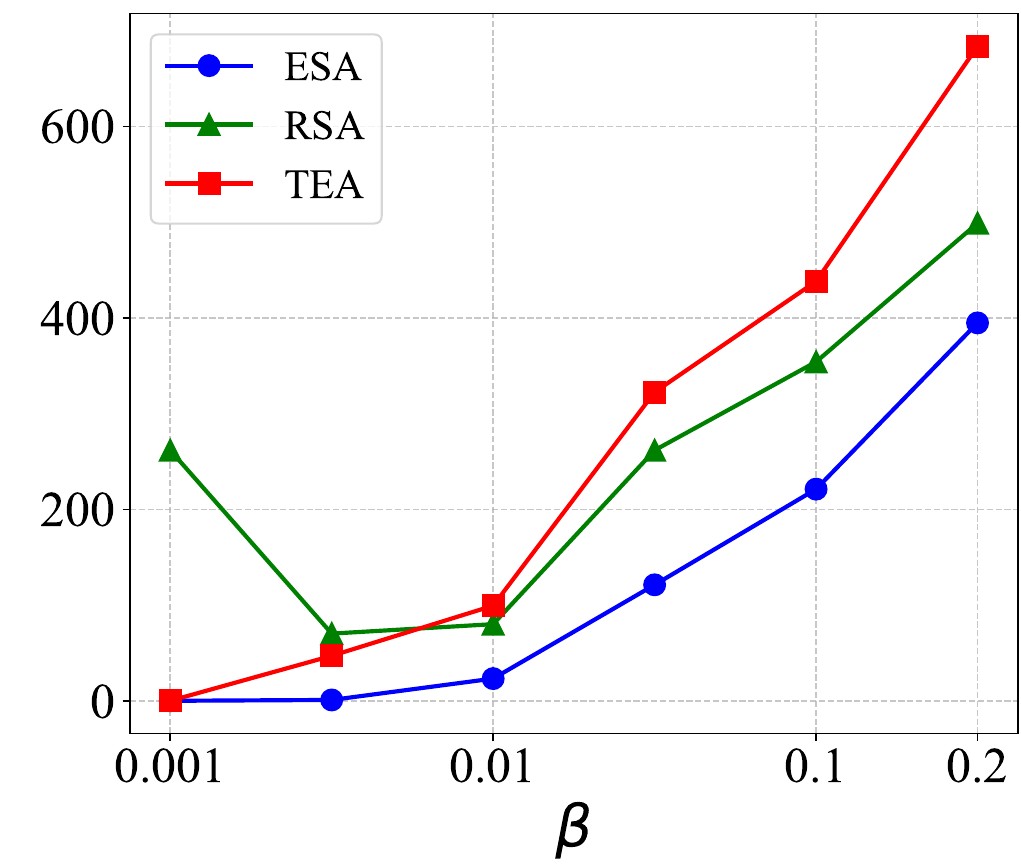}
  & \includegraphics[width=\linewidth]{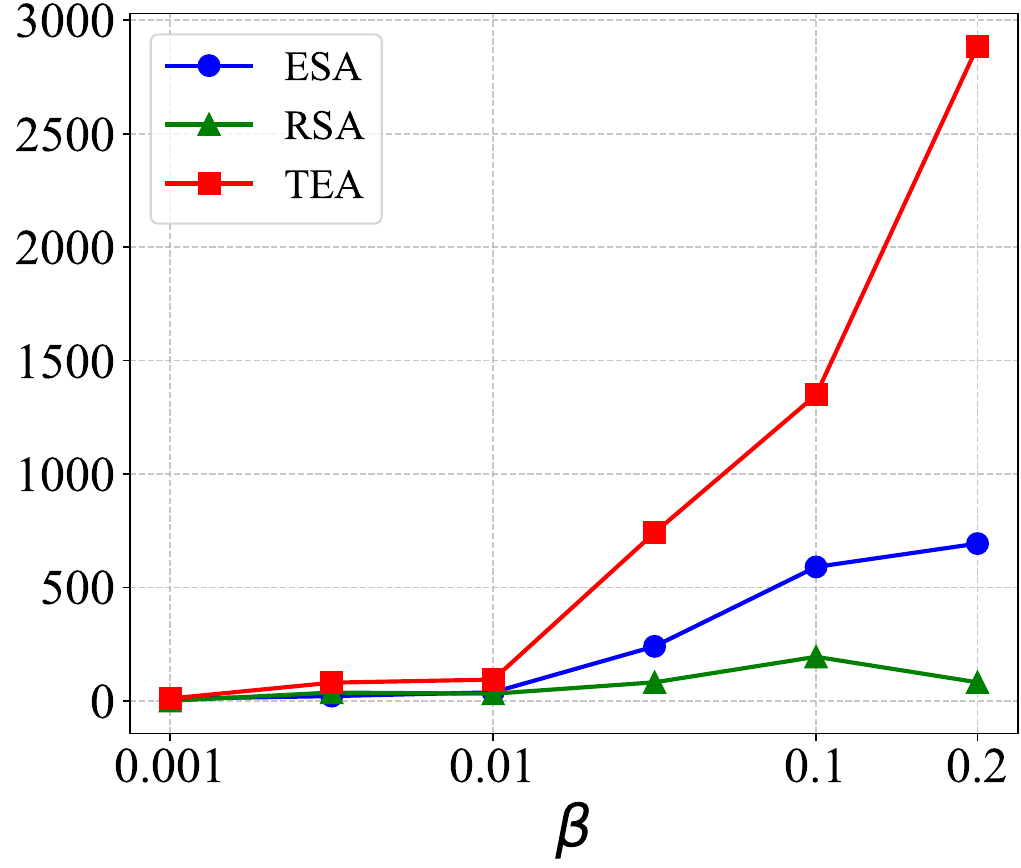}
  & \includegraphics[width=\linewidth]{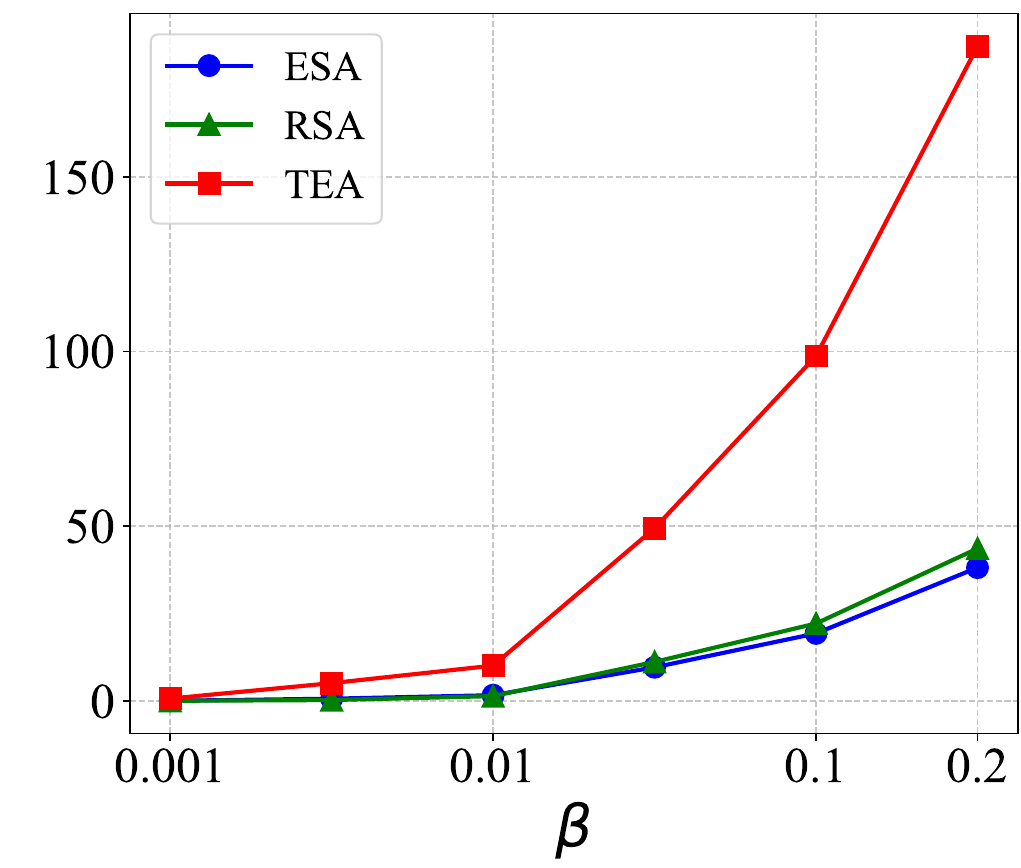} \\[-2pt]

  \rotatebox[origin=c]{90}{\normalfont\scriptsize Gain (TC)}
  & \includegraphics[width=\linewidth]{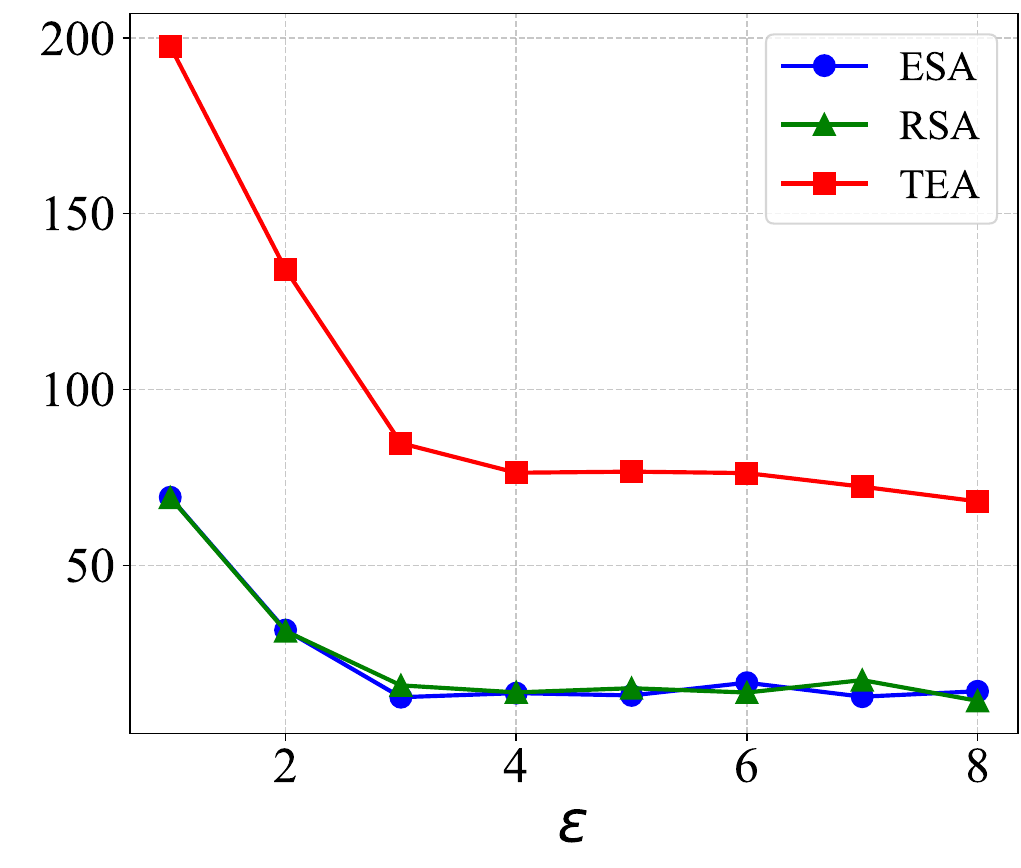}
  & \includegraphics[width=\linewidth]{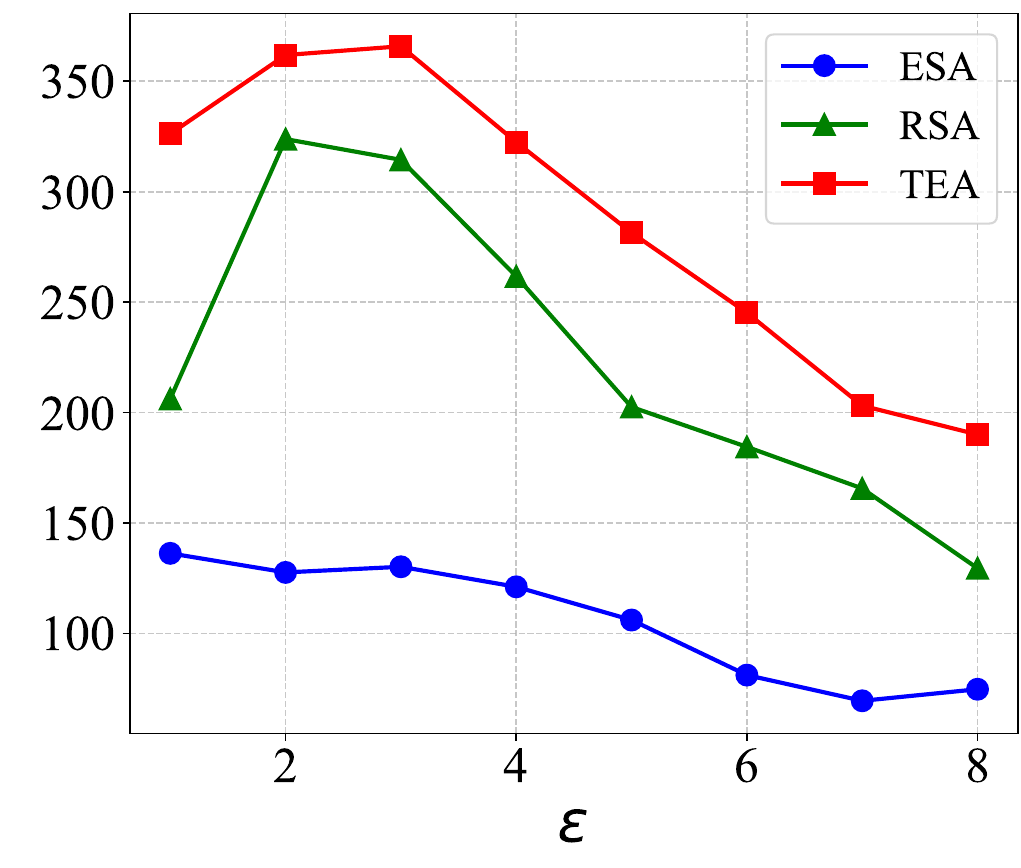}
  & \includegraphics[width=\linewidth]{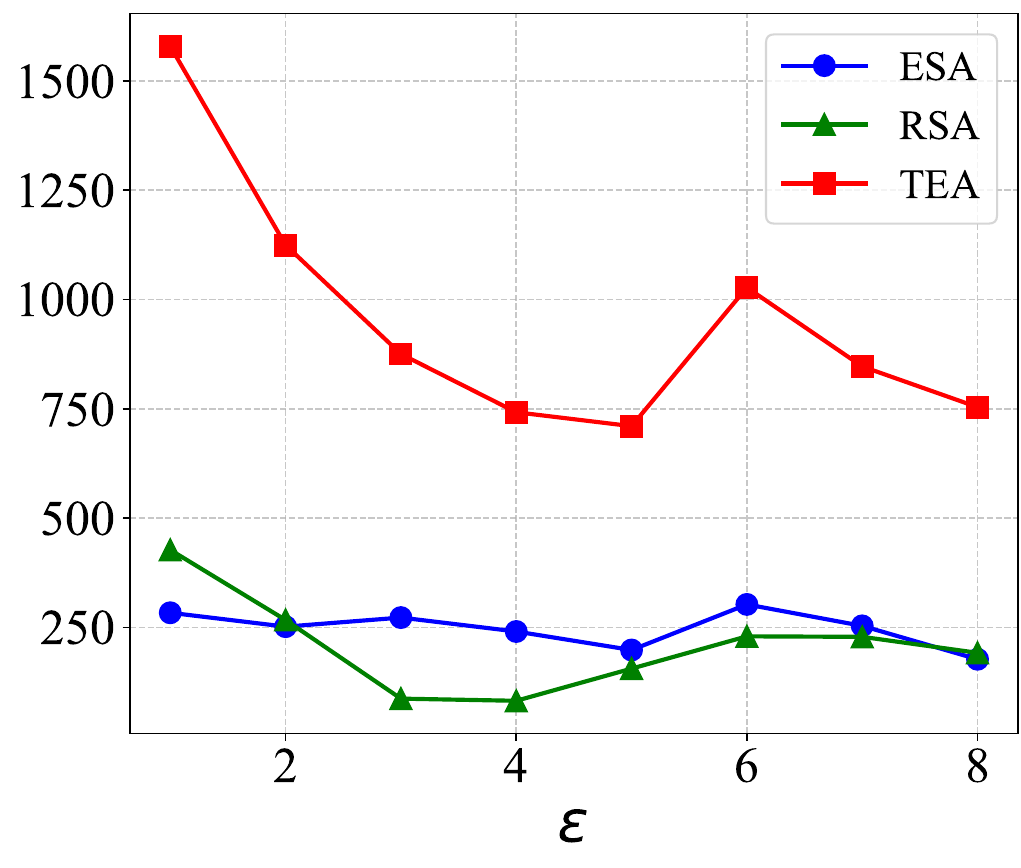}
  & \includegraphics[width=\linewidth]{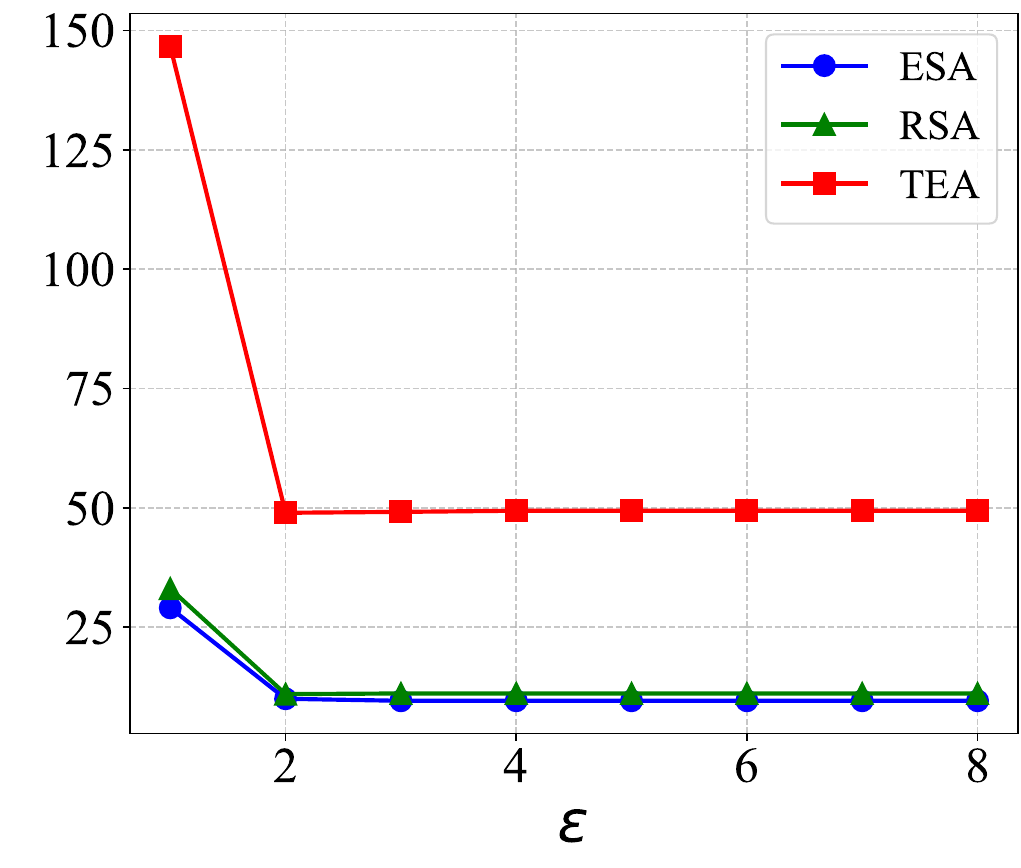} \\[-2pt]

  \rotatebox[origin=c]{90}{\normalfont\scriptsize Gain (TC)}
  & \includegraphics[width=\linewidth]{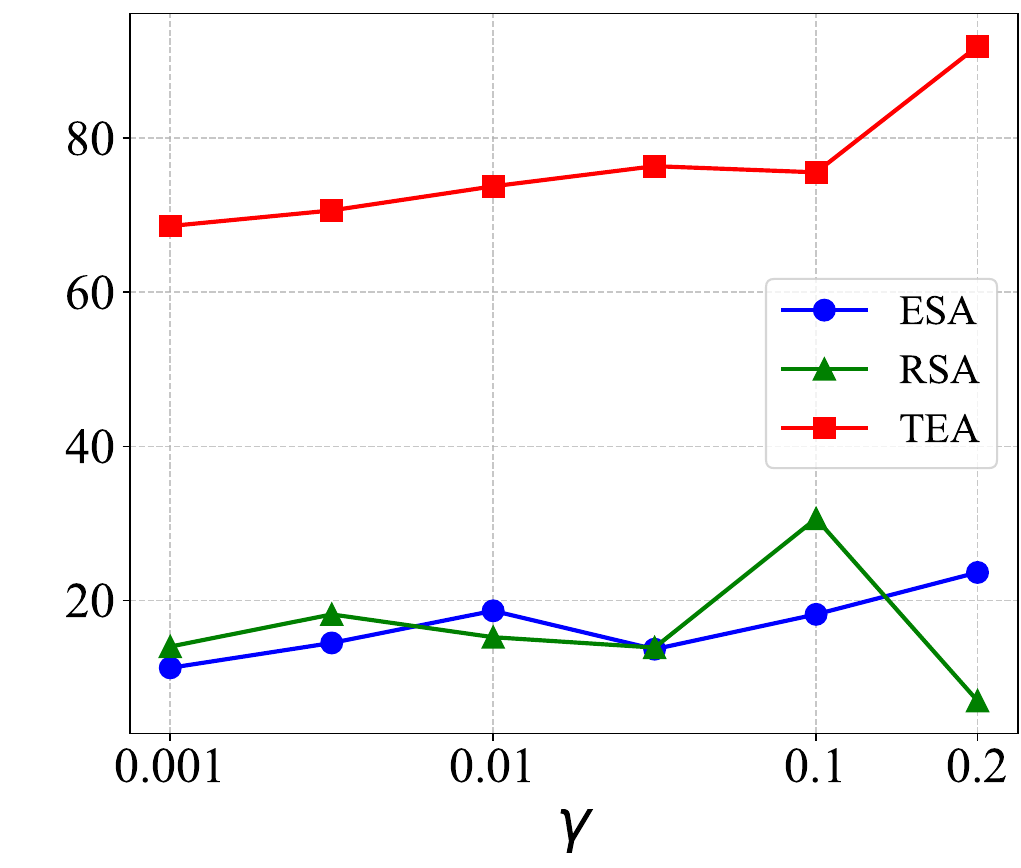}
  & \includegraphics[width=\linewidth]{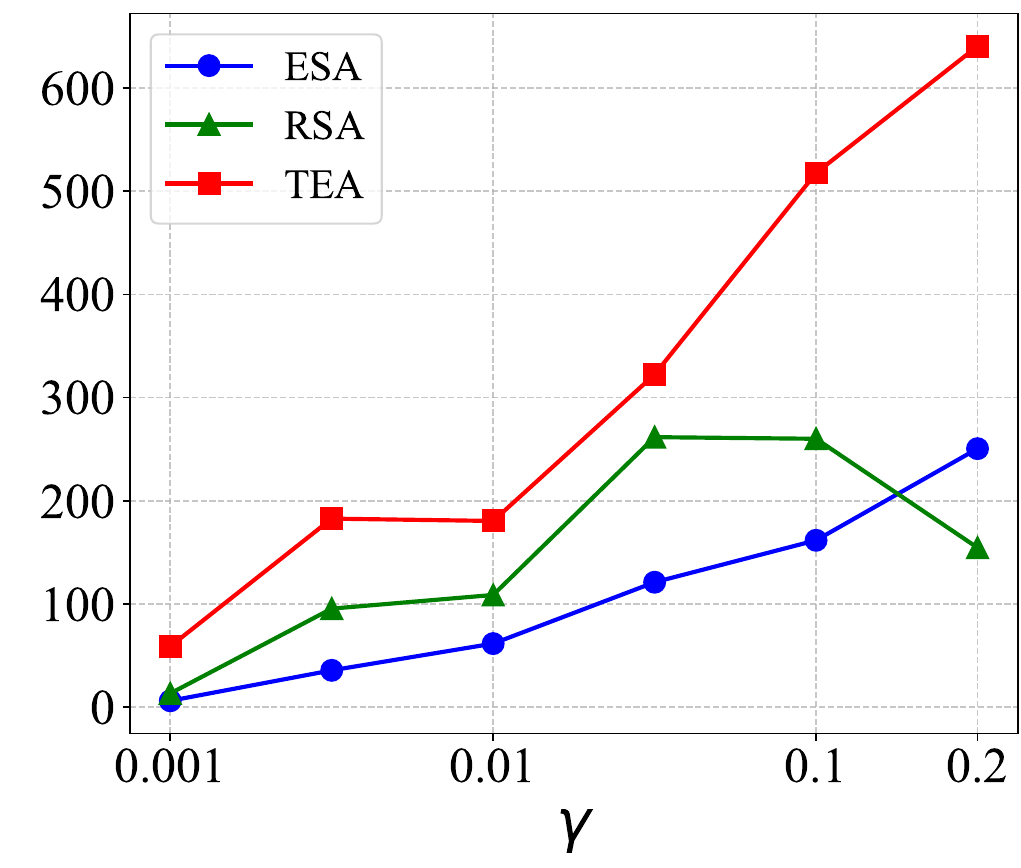}
  & \includegraphics[width=\linewidth]{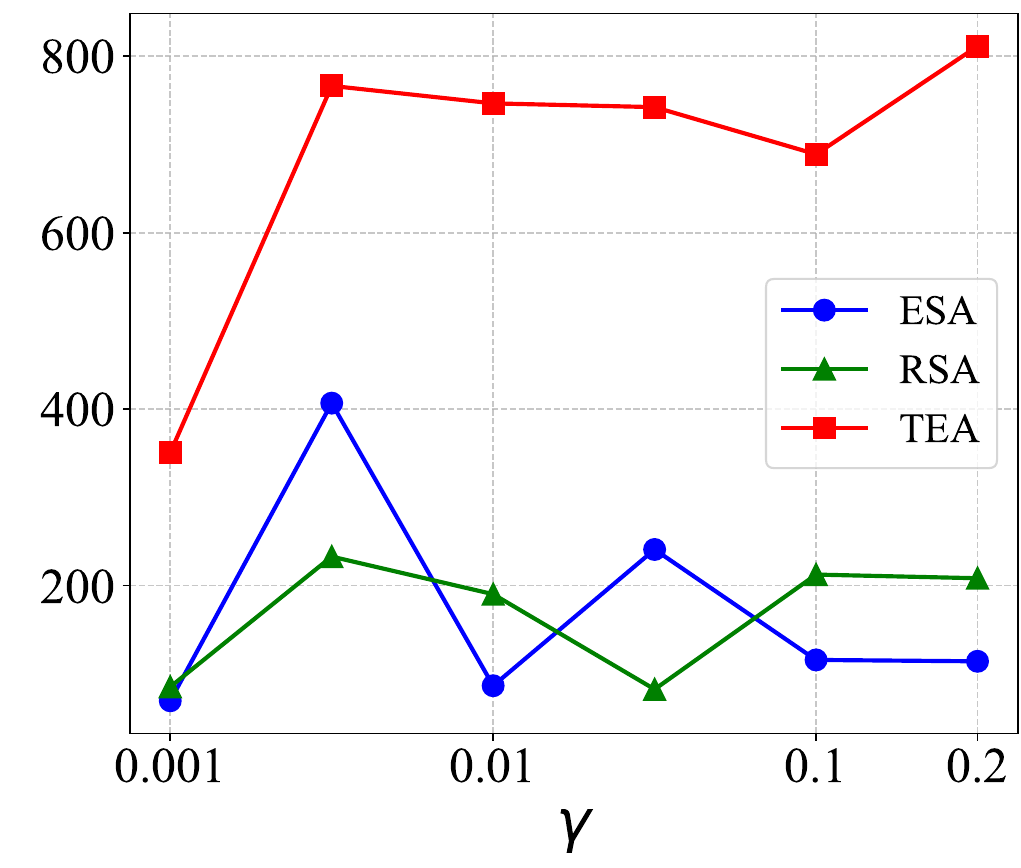}
  & \includegraphics[width=\linewidth]{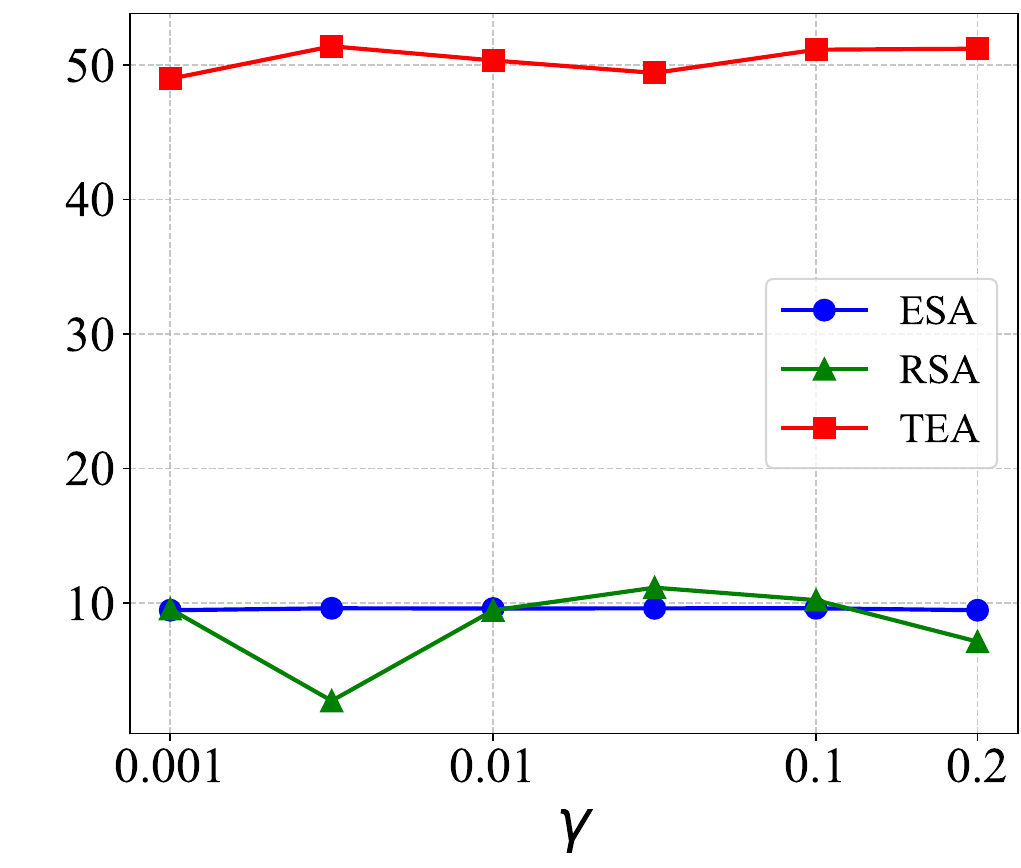} \\

\end{tabular}%
}
\caption{The effect of $\beta$, $\epsilon$ and $\gamma$ changes on gain of triangle count}
\label{fig:trianglecount_mga}
\end{figure*}

\textbf{Privacy Budget $\epsilon$:} The second row illustrates the effect of parameter $\epsilon$ on the gain of degree centrality. Across all four datasets, a consistent decreasing trend is observed in the low $\epsilon$ range (approximately 1.0–4.0) as $\epsilon$ increases. This can be explained as follows: in the proposed framework, the edge injection budget of MGA is determined by the average degree upper bound $\bar{d}_{t_i}$ estimated from the HMG-reconstructed graph. When $\epsilon$ is small, strong LDP noise introduces a large number of noisy edges, leading to an overestimation of this upper bound and thus amplifying the attack capability of fake nodes. As $\epsilon$ increases, the noise weakens, the upper bound decreases, and the attack space becomes constrained, resulting in a reduction in gain.

\begin{figure*}[t]
\setlength{\tabcolsep}{1.5pt} 
\resizebox{\textwidth}{!}{%
\begin{tabular}{@{} >{\centering\arraybackslash}m{0.3cm} @{\hspace{1pt}}
                >{\centering\arraybackslash}m{0.245\textwidth}
                >{\centering\arraybackslash}m{0.245\textwidth}
                >{\centering\arraybackslash}m{0.245\textwidth}
                >{\centering\arraybackslash}m{0.245\textwidth} @{}}

  & {\normalfont\scriptsize Ass-733}
  & {\normalfont\scriptsize Email}
  & {\normalfont\scriptsize Reply}
  & {\normalfont\scriptsize Synthetic} \\[1pt]

  \rotatebox[origin=c]{90}{\normalfont\scriptsize Gain (CC)}
  & \includegraphics[width=\linewidth]{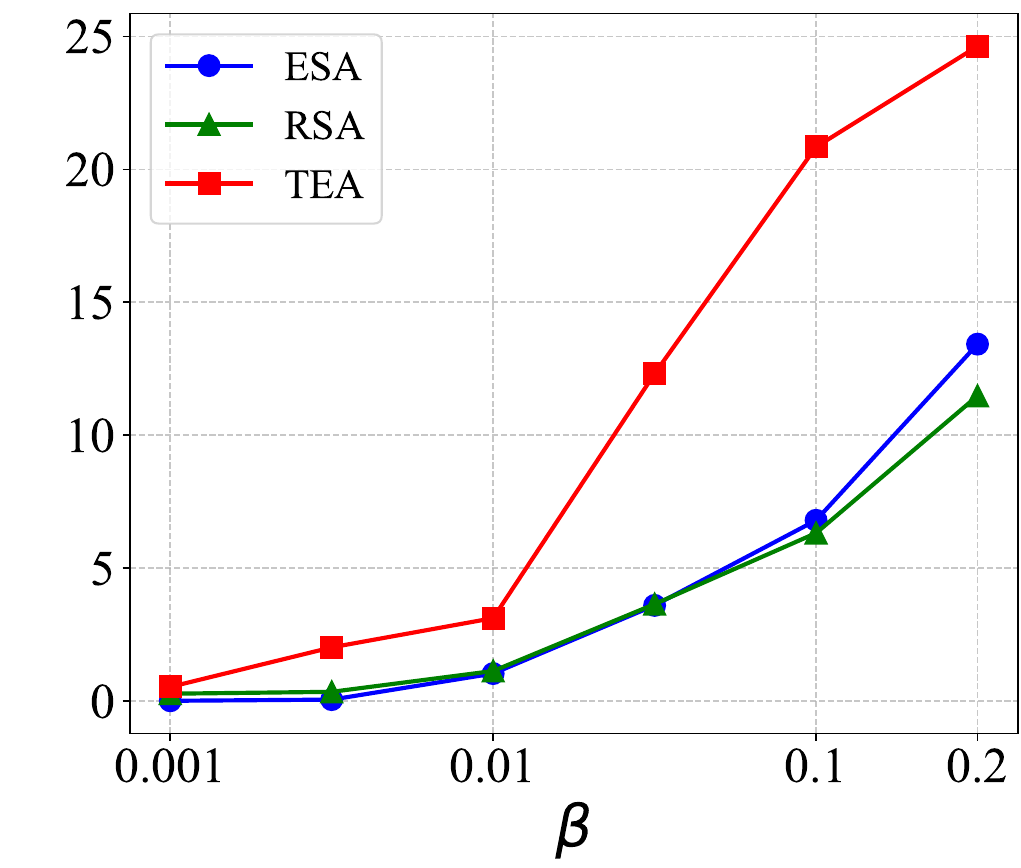}
  & \includegraphics[width=\linewidth]{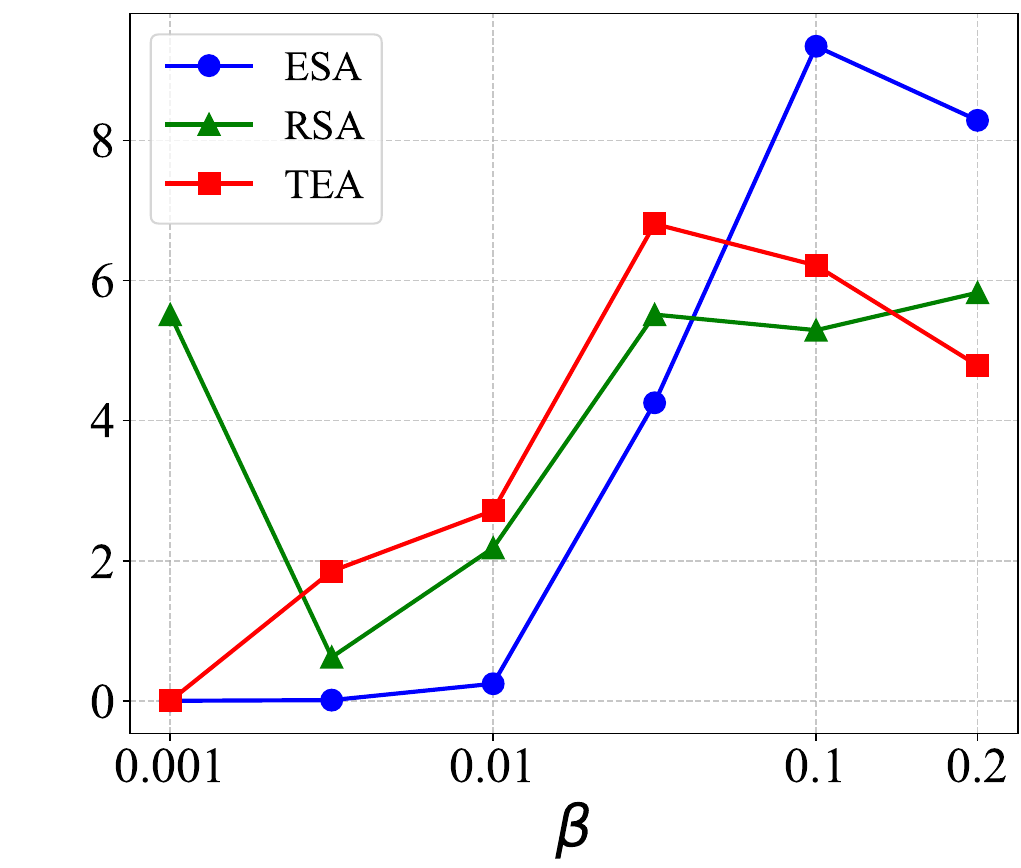}
  & \includegraphics[width=\linewidth]{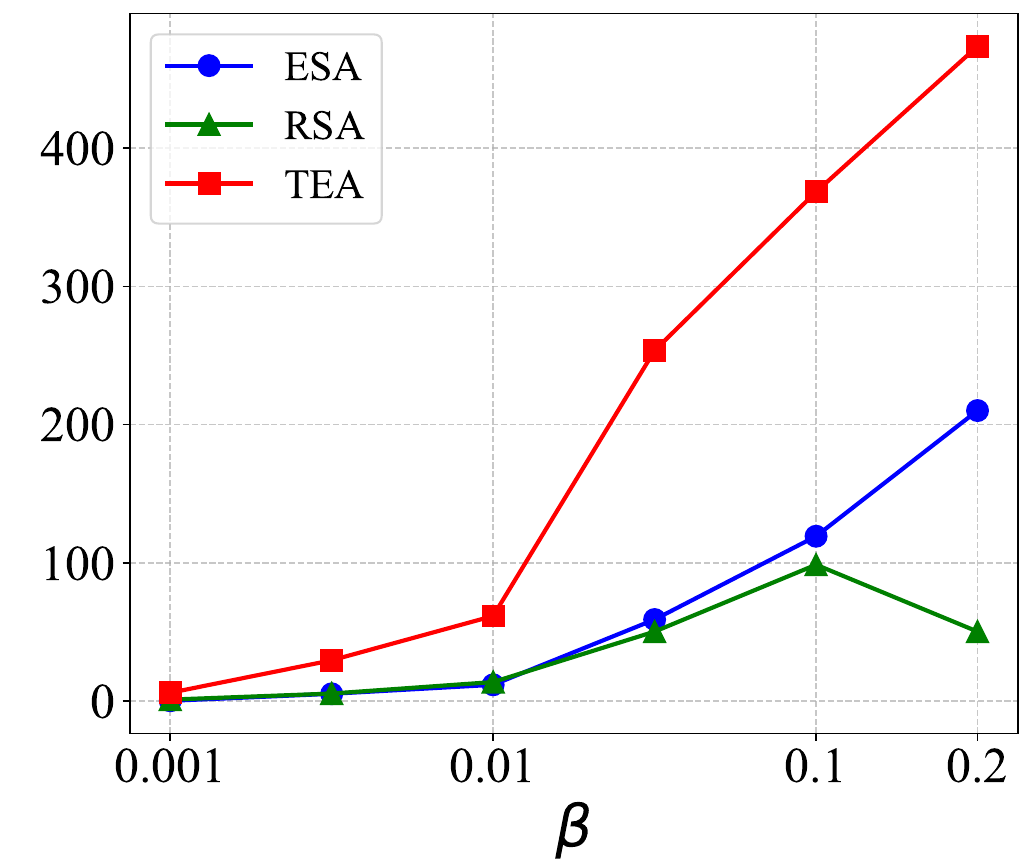}
  & \includegraphics[width=\linewidth]{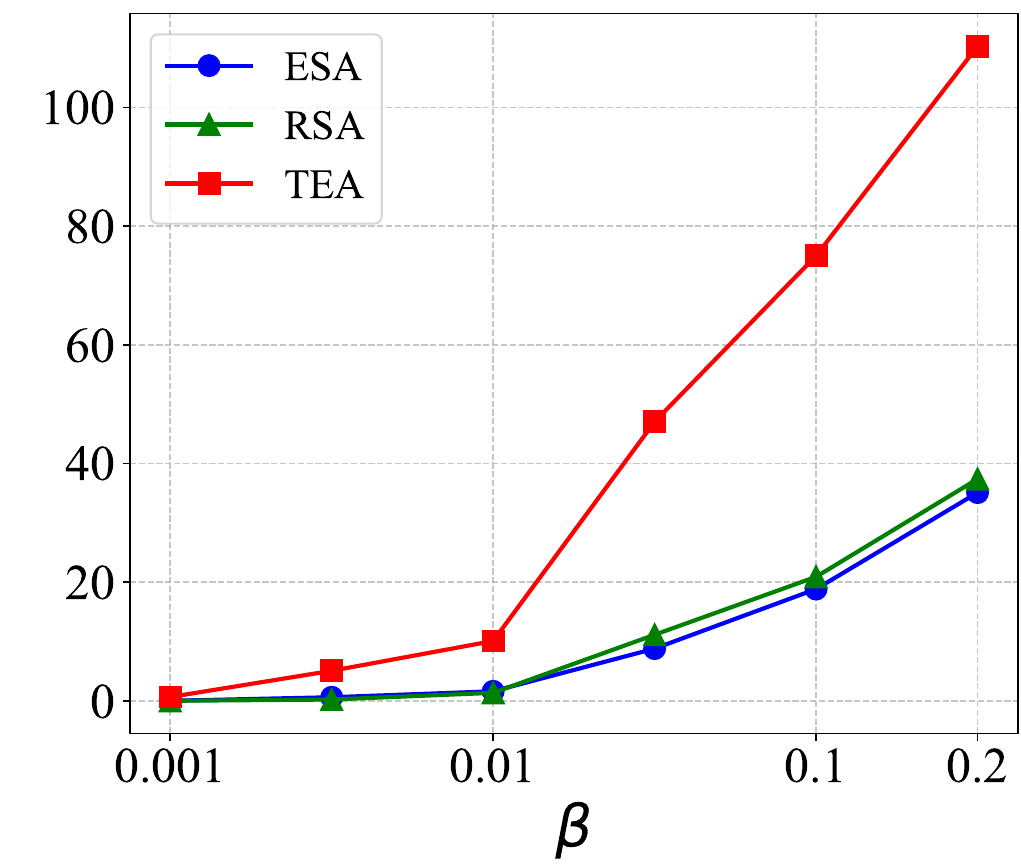} \\[-2pt]

  \rotatebox[origin=c]{90}{\normalfont\scriptsize Gain (CC)}
  & \includegraphics[width=\linewidth]{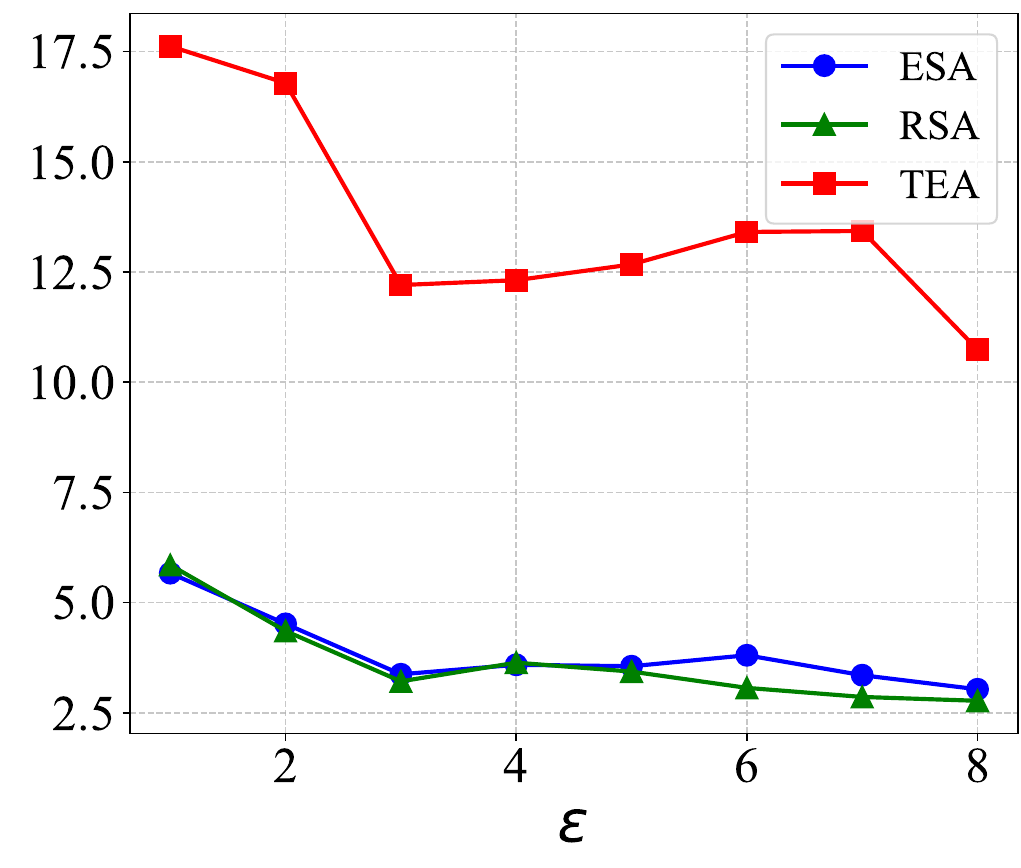}
  & \includegraphics[width=\linewidth]{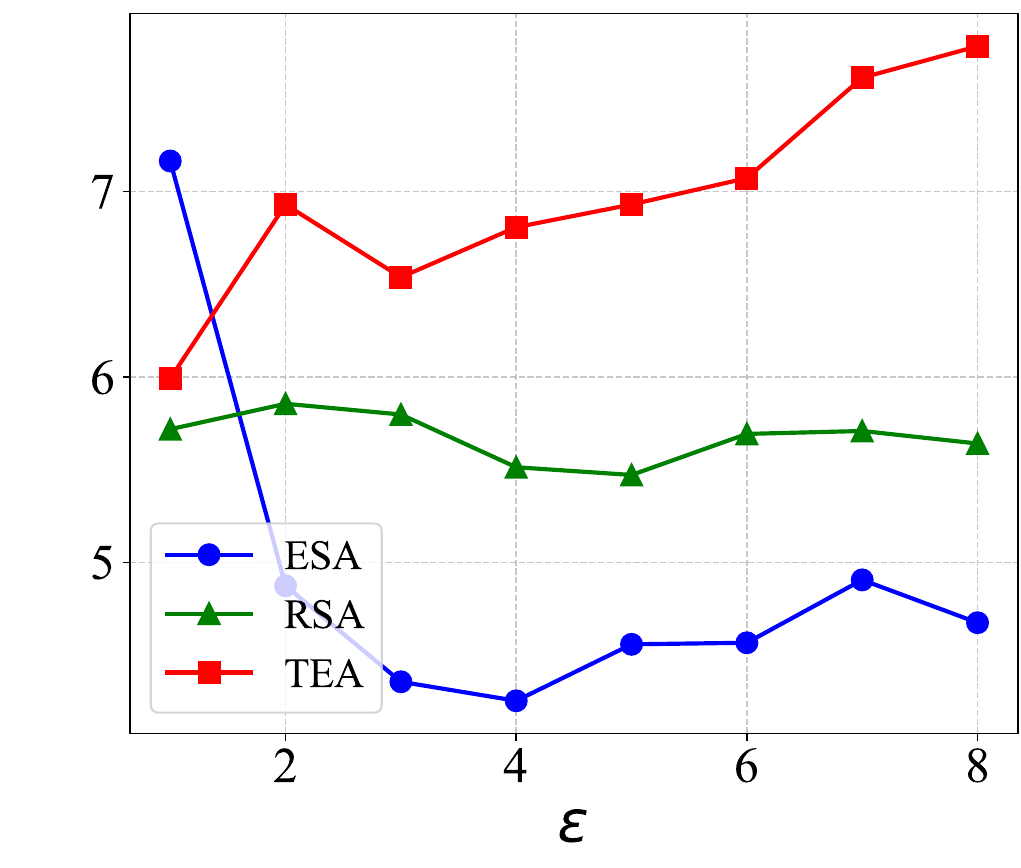}
  & \includegraphics[width=\linewidth]{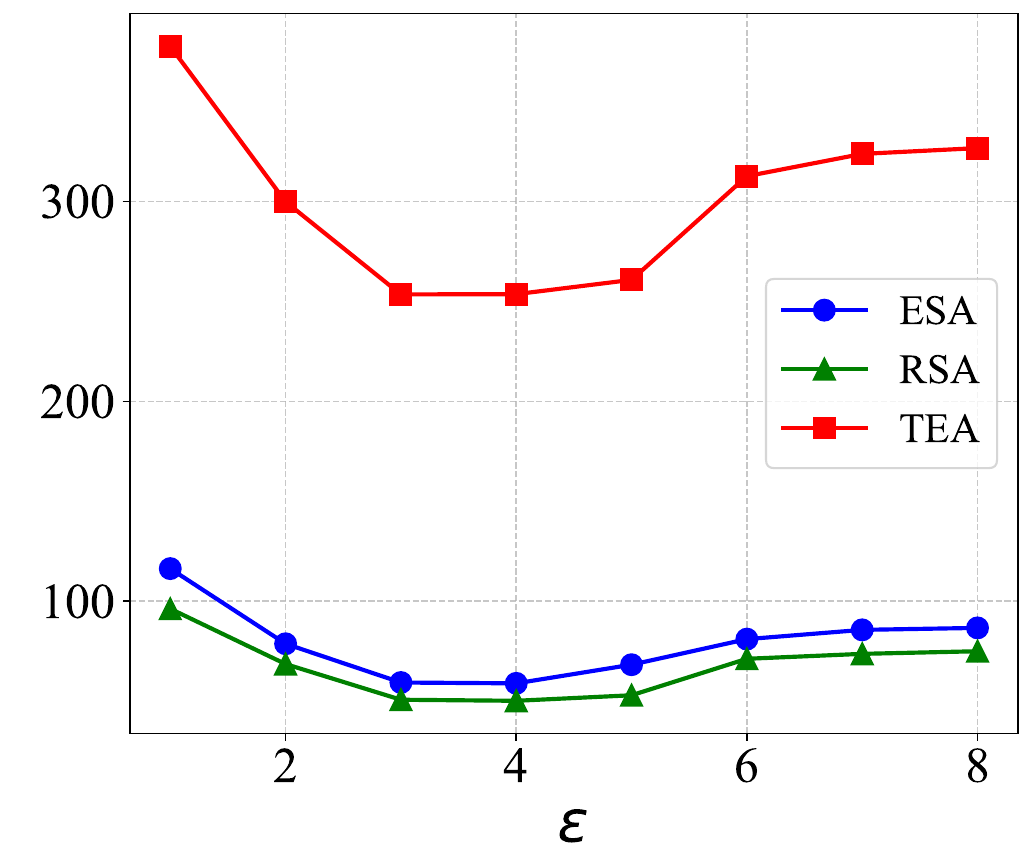}
  & \includegraphics[width=\linewidth]{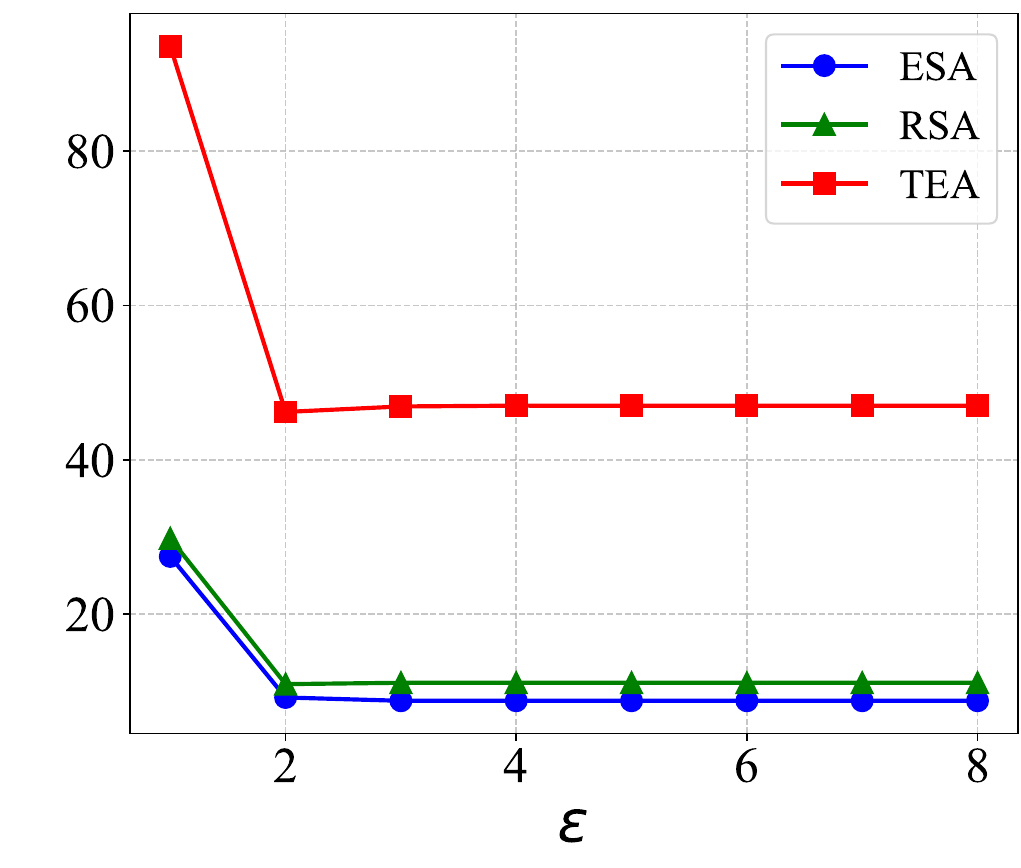} \\[-2pt]

  \rotatebox[origin=c]{90}{\normalfont\scriptsize Gain (CC)}
  & \includegraphics[width=\linewidth]{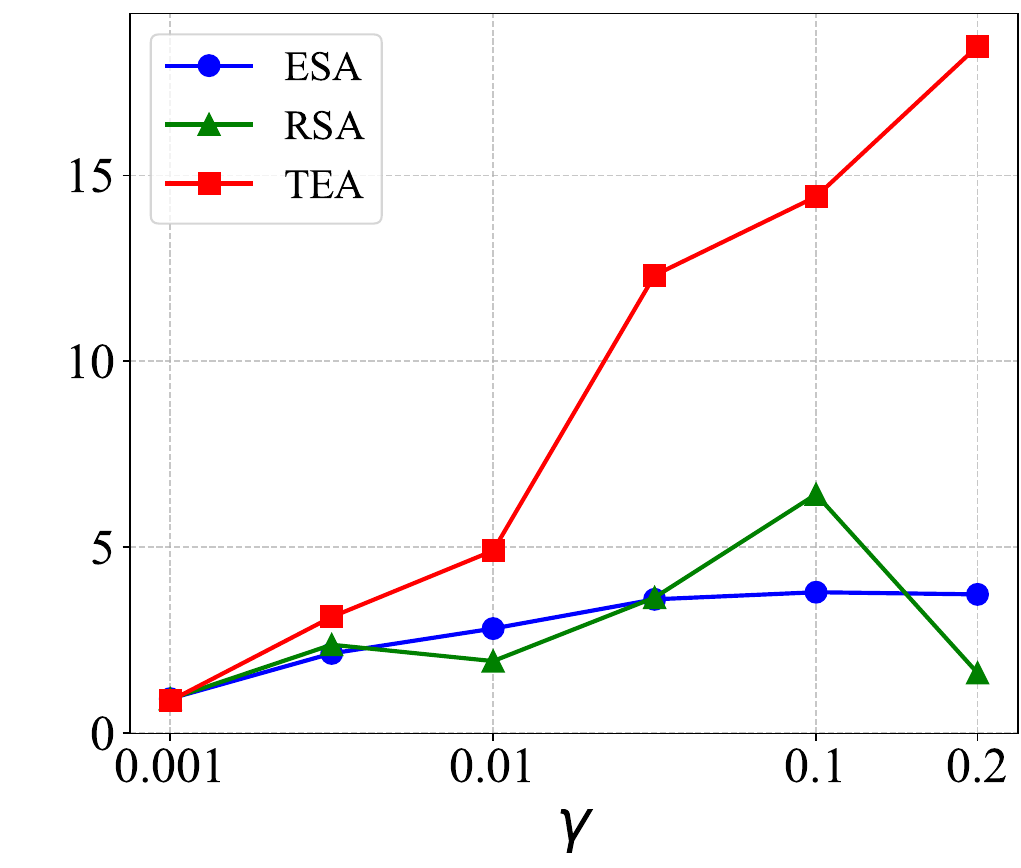}
  & \includegraphics[width=\linewidth]{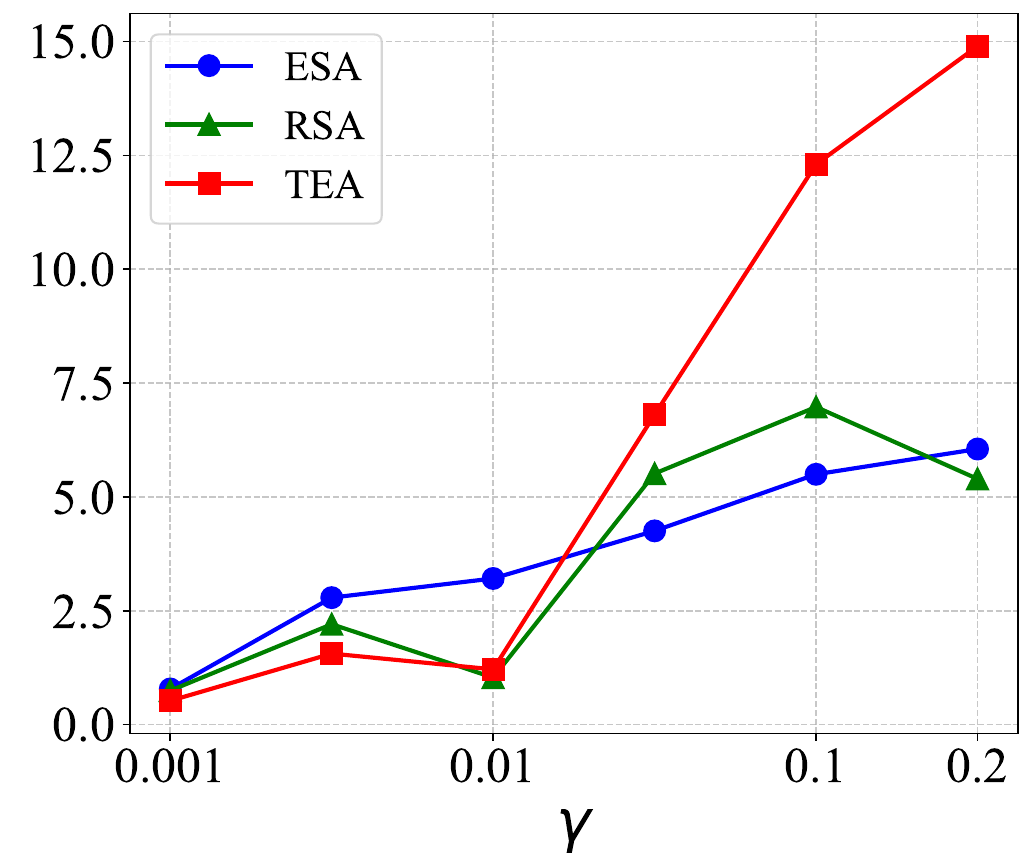}
  & \includegraphics[width=\linewidth]{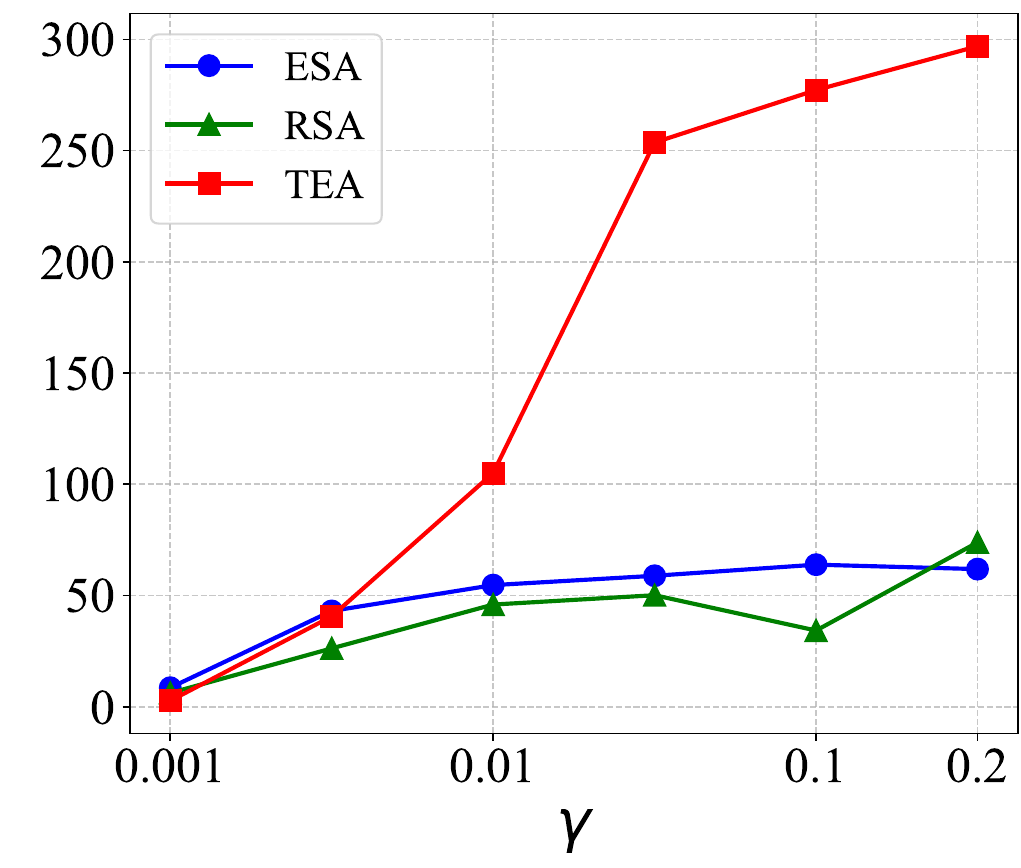}
  & \includegraphics[width=\linewidth]{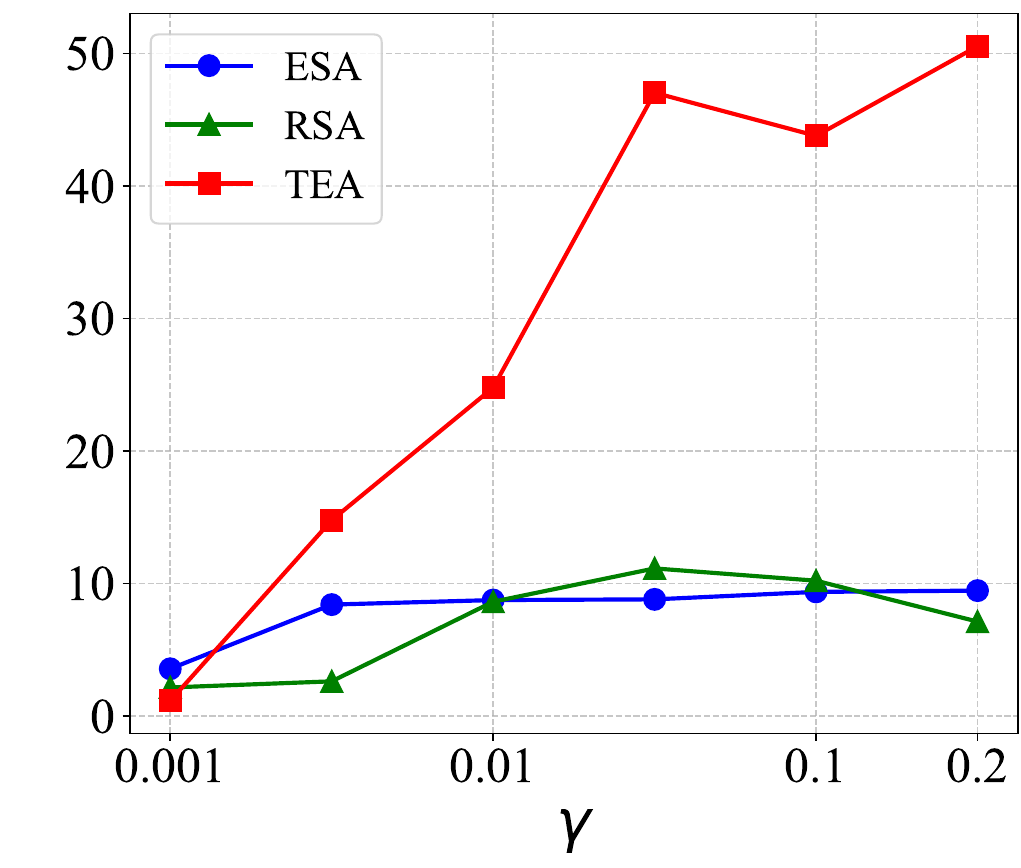} \\

\end{tabular}%
}
\caption{The effect of $\beta$, $\epsilon$ and $\gamma$ changes on gain of clustering coefficient}
\label{fig:gamma_clustering_mga}
\end{figure*}

Among the datasets, the Email dataset exhibits a slight rebound as $\epsilon$ increases, which can be attributed to the combined effects of temporal heterogeneity and weighted aggregation. In contrast, the Reply dataset demonstrates a “decrease-then-increase” pattern, primarily due to its large-scale sparsity. Specifically, as $\epsilon$ increases, the number of active nodes declines faster than the number of edges, thereby raising the upper bound of the average degree, enlarging the edge allocation budget, and ultimately leading to a recovery in gain. By comparison, the remaining datasets maintain a relatively stable proportion of active nodes, with only mild variations in average degree, and thus do not exhibit a similar rebound.

\textbf{Target Node Ratio $\gamma$:} The third row illustrates the impact of the target node ratio $\gamma$ on the degree centrality gain. A common pattern across all four datasets is that TEA consistently outperforms ESA and RSA over the entire range, and the gain generally increases at higher $\gamma$ values (approximately 0.1--0.2). This is because increasing $\gamma$ expands the target node set, thereby increasing the summation term, while also raising the probability that injected edges connect to target nodes. In the small-to-moderate $\gamma$ regime, the total attack budget is constrained by a fixed $\beta$, leading to its allocation being distributed across a larger number of target nodes. As a result, ESA and RSA generally remain relatively stable or exhibit a slight decreasing trend.

\subsection{Results of Triangle Count}

Figure \ref{fig:trianglecount_mga} illustrates the impact of varying parameters $\beta$, $\epsilon$, and $\gamma$ on the triangle count gain. Due to the high sensitivity of triangle count to local structures, newly added edges often produce multiplicative amplification effects, resulting in pronounced nonlinear growth or abrupt transitions in the curves.

\textbf{Fake node ratio $\beta$:} The first row corresponding to variations in $\beta$ exhibits a consistent increasing trend in gain as $\beta$ grows, with TEA showing the most significant improvement. This indicates that under expanded attack budgets, large-scale concentrated attacks are more effective in continuously triggering triangle closure. In the Reply dataset, TEA achieves the most pronounced increase, suggesting that its local structures are more amenable to closure and can be more effectively exploited in weighted time slices.

\textbf{Privacy budget $\epsilon$:} The second row ($\varepsilon$ variation) exhibits a general “decrease-then-stabilization or slight fluctuation” pattern, indicating that as $\varepsilon$ increases, the attack budget contracts, reducing the number of effective edges available for triangle closure and leading to a decline and eventual stabilization of the gain. The Reply dataset shows a rebound around $\varepsilon = 6.0$, suggesting a phase-wise expansion of the feasible edge set in this range; the Synthetic dataset stabilizes for $\varepsilon \geq 2.0$, indicating rapid saturation of its triangle closure space; in the Email dataset, TEA and RSA exhibit local peaks in the moderate $\varepsilon$ regime followed by a decline, reflecting a trade-off between higher closure efficiency under moderate constraints and diminishing marginal gains as constraints tighten.

\textbf{Target node ratio $\gamma$:} The third row ($\gamma$ variation) shows that as $\gamma$ increases, TEA maintains a relatively high level and exhibits an overall upward trend, ESA generally increases gradually or remains stable, while RSA displays more pronounced fluctuations. This is mainly because enlarging the target set introduces more potential opportunities for triangle closure, and TEA is more capable of concentrating its limited budget on high-weight time slices. In the Reply dataset, ESA and RSA exhibit a distinct “zigzag” pattern, indicating high sensitivity to local critical structures; in contrast, the Synthetic dataset shows relatively minor variations across all three strategies, suggesting that target expansion has a limited impact on its triangle closure opportunities.

\subsection{Results of Clustering Coefficient}

Figure \ref{fig:gamma_clustering_mga}  illustrates the impact of varying parameters $\beta$, $\epsilon$, and $\gamma$ on the gain of clustering coefficient of target nodes. Overall, the relative performance among the three attack strategies remains consistent, with TEA generally outperforming ESA and RSA. Compared to degree centrality, the clustering coefficient relies more heavily on local triadic structures, making it more susceptible to nonlinear variations and phase-wise turning points.

\textbf{Fake node ratio $\beta$:} The first row corresponding to variations in $\beta$ exhibits a consistent trend: as $\beta$ increases, the gain improves overall, with TEA showing the most significant growth. This is because $\beta$ directly determines the scale of fake nodes and the budget of injectable edges, and a larger budget facilitates stronger perturbations of local triadic structures around the target nodes. 

Notably, in the Email and Reply datasets, RSA shows a decline in the high-$\beta$ regime, and a similar phenomenon is observed for TEA in Email. This is mainly due to the saturation of local structures around target nodes and the constraints on the number of edges that fake nodes can establish. In addition, the random-scale scheduling introduces temporal fluctuations in the budget across time slices, further weakening its effectiveness. In contrast, AS-733 and Synthetic exhibit smoother growth, indicating that their structures respond more stably to budget expansion.

\textbf{Privacy budget $\epsilon$:} The second row ($\varepsilon$ variation) exhibits a consistent overall pattern: most curves initially decrease with increasing $\varepsilon$ in the low-to-moderate regime, and then transition to a stable or slowly varying phase. This behavior is mainly due to the attack budget upper bound being constrained by the HMG reconstruction statistics; as $\varepsilon$ increases, reduced noise compresses the available attack space. Differences emerge across datasets in the mid-to-high $\varepsilon$ range: the Reply dataset shows a “decrease-then-increase” pattern, driven by its sparsity, which causes the statistical upper bound to rebound at high $\varepsilon$, thereby enlarging the edge injection budget; in the Email dataset, TEA continues to increase, as higher-weight time slices facilitate triangle closure; whereas the Synthetic dataset rapidly declines and then stabilizes, indicating limited variation in its statistical upper bound at higher $\varepsilon$.

\begin{figure*}[t]
\setlength{\tabcolsep}{1.5pt} 
\resizebox{\textwidth}{!}{%
\begin{tabular}{@{} >{\centering\arraybackslash}m{0.3cm} @{\hspace{1pt}}
                >{\centering\arraybackslash}m{0.245\textwidth}
                >{\centering\arraybackslash}m{0.245\textwidth}
                >{\centering\arraybackslash}m{0.245\textwidth}
                >{\centering\arraybackslash}m{0.245\textwidth} @{}}

  & {\normalfont\scriptsize Ass-733}
  & {\normalfont\scriptsize Email}
  & {\normalfont\scriptsize Reply}
  & {\normalfont\scriptsize Synthetic} \\[1pt]

  \rotatebox[origin=c]{90}{\normalfont\scriptsize Gain (ASPL)}
  & \includegraphics[width=\linewidth]{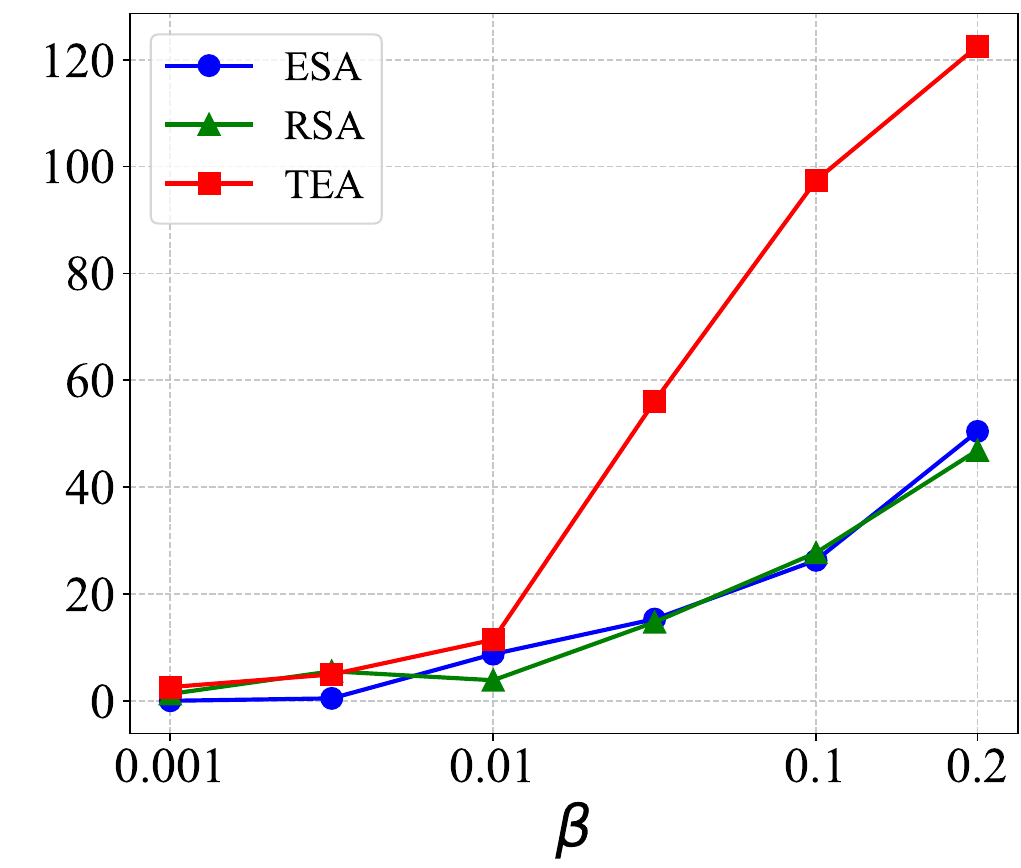}
  & \includegraphics[width=\linewidth]{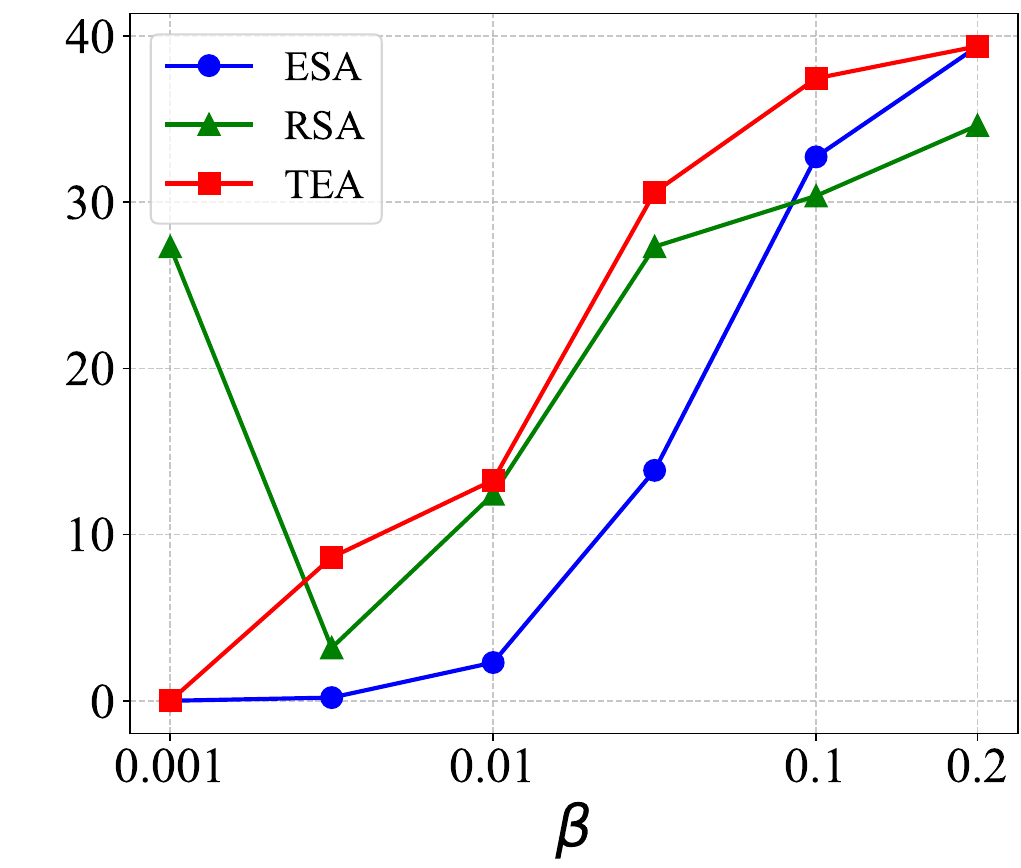}
  & \includegraphics[width=\linewidth]{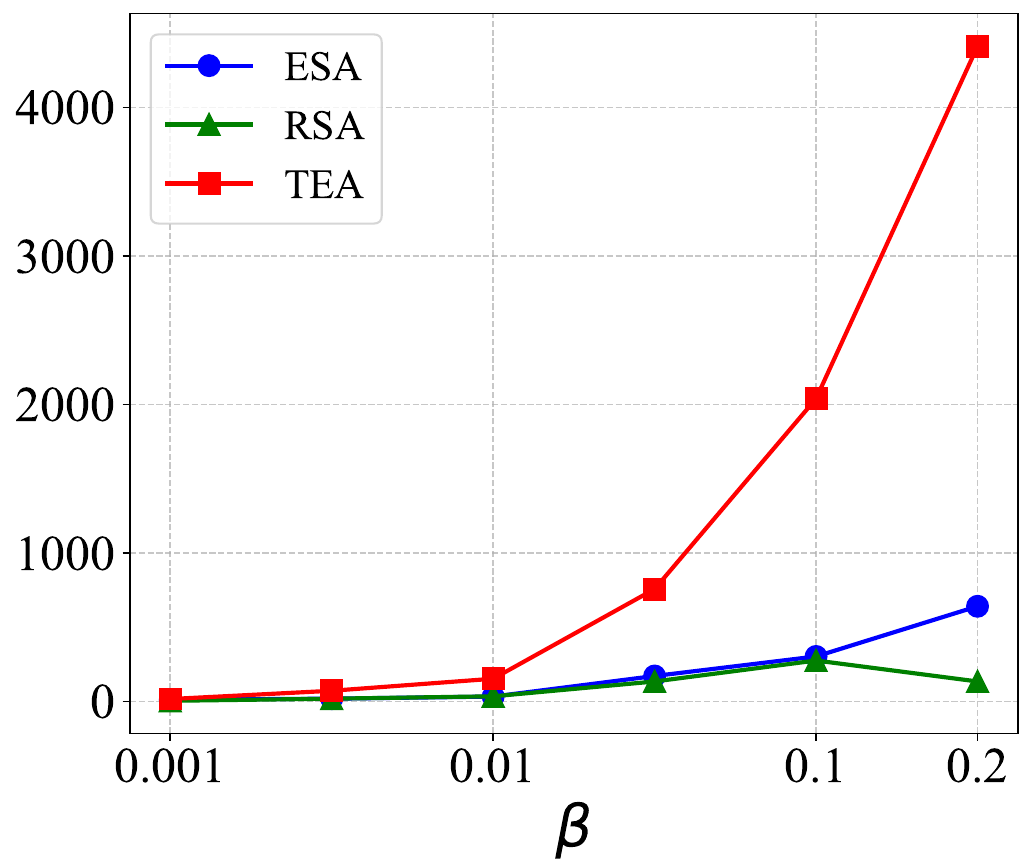}
  & \includegraphics[width=\linewidth]{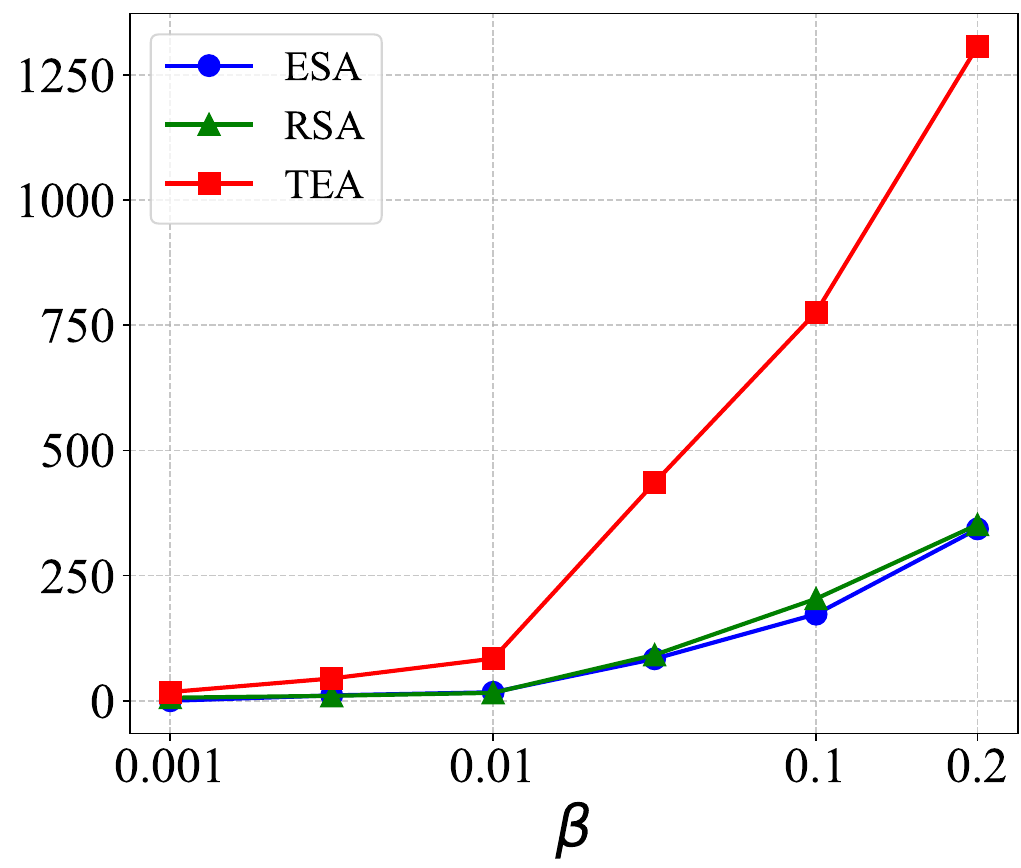} \\[-2pt]

  \rotatebox[origin=c]{90}{\normalfont\scriptsize Gain (ASPL)}
  & \includegraphics[width=\linewidth]{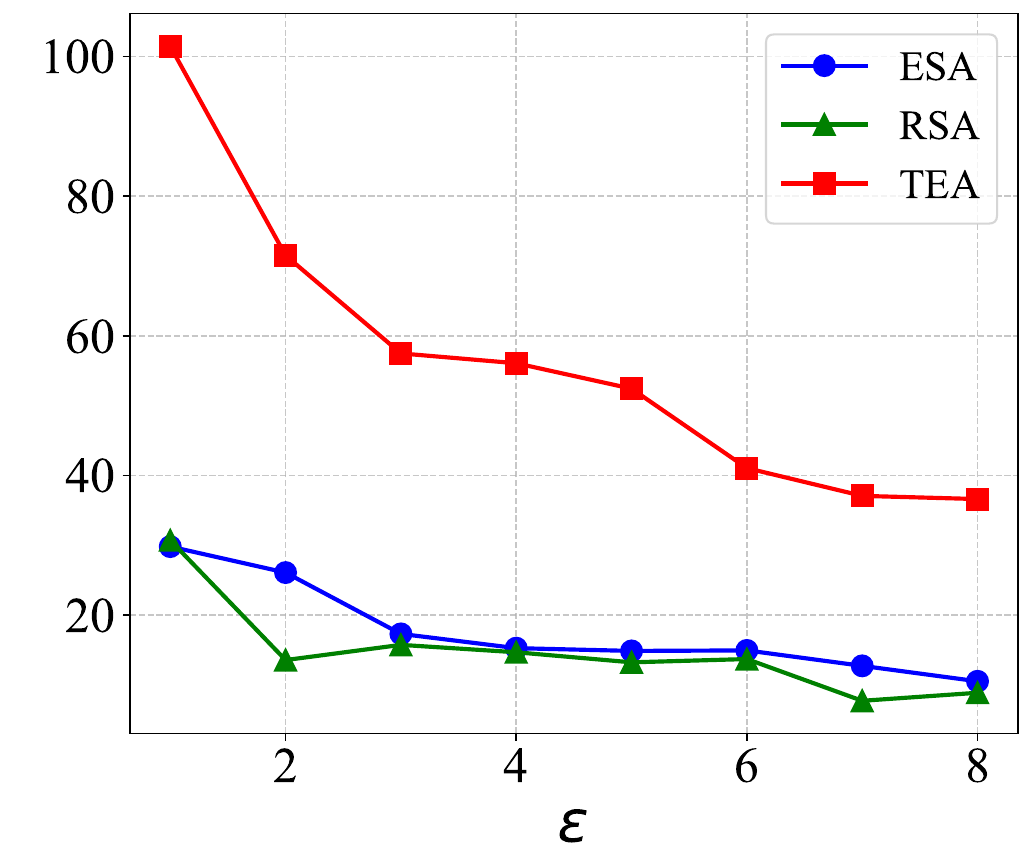}
  & \includegraphics[width=\linewidth]{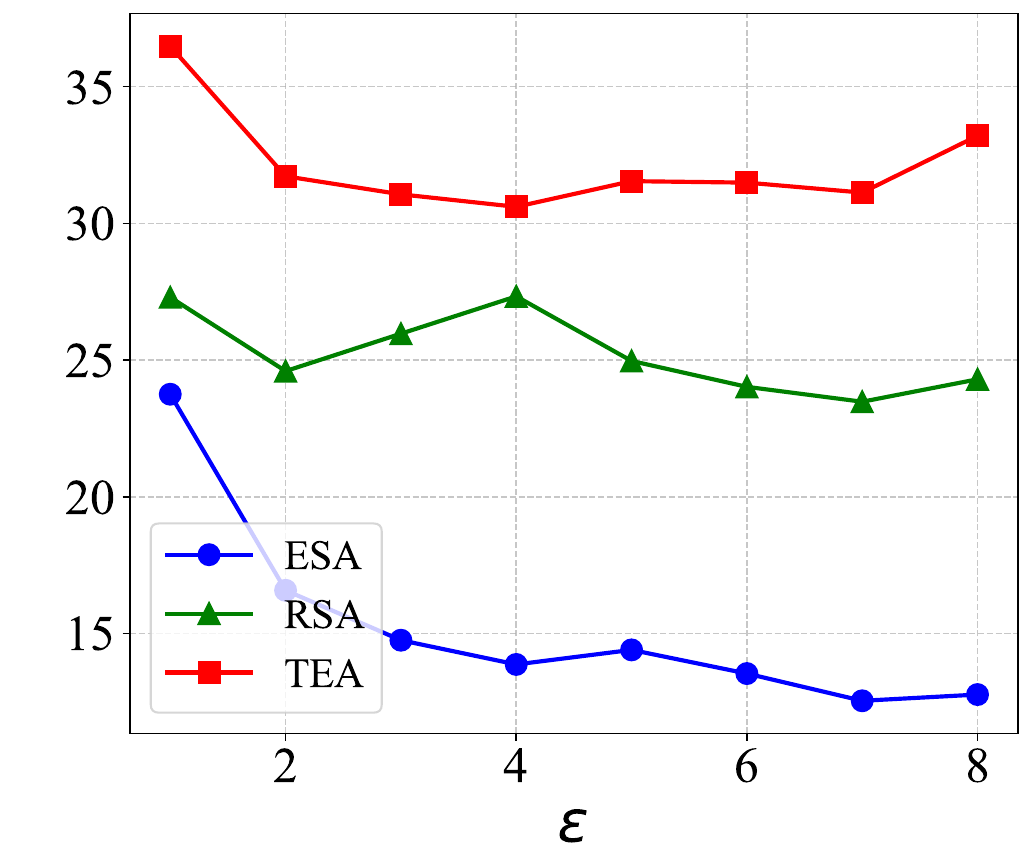}
  & \includegraphics[width=\linewidth]{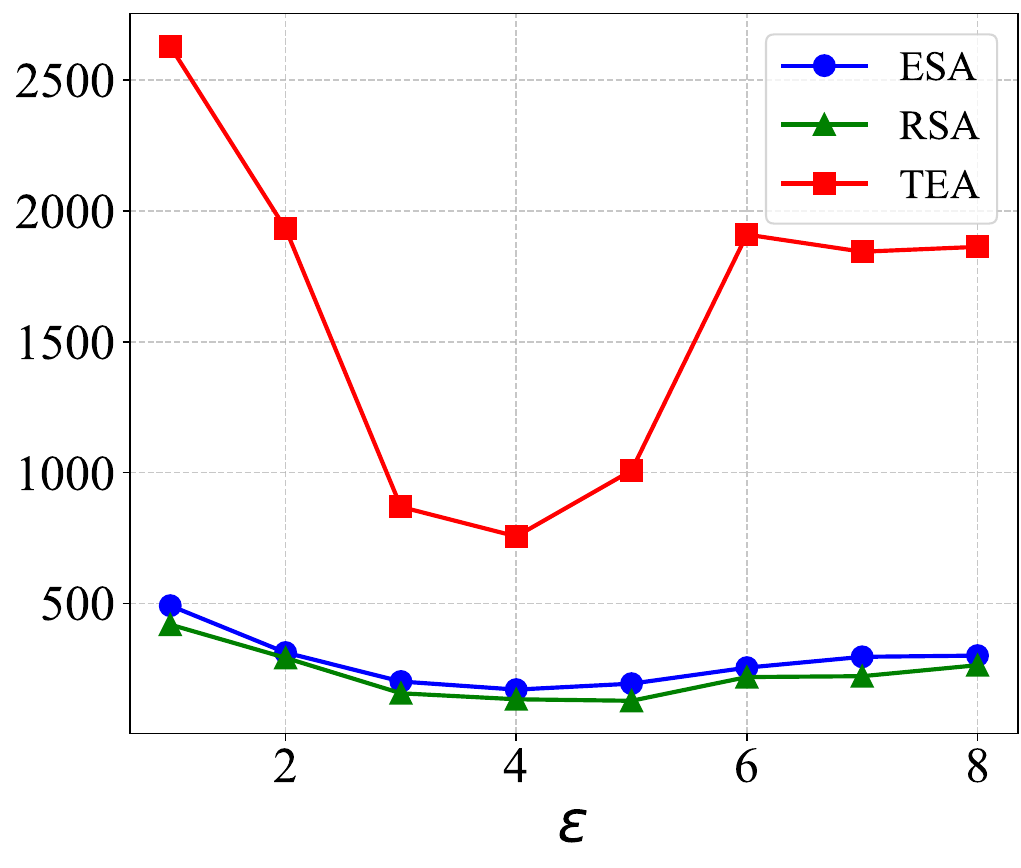}
  & \includegraphics[width=\linewidth]{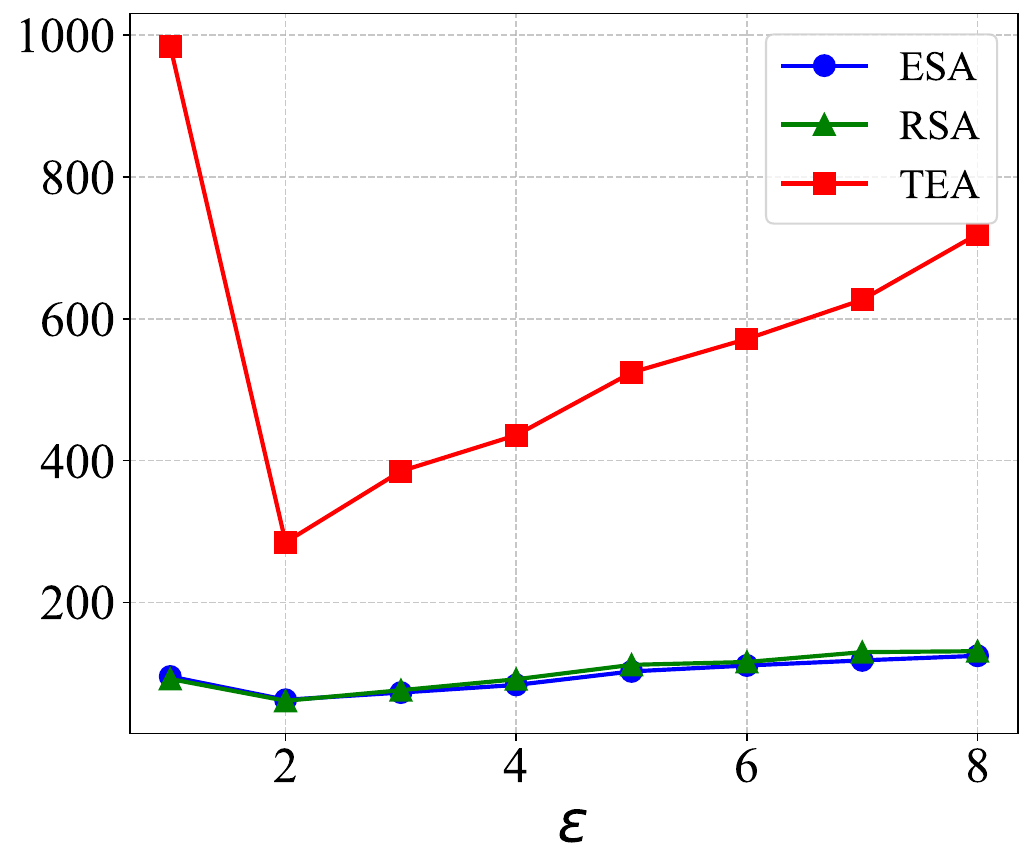} \\[-2pt]

  \rotatebox[origin=c]{90}{\normalfont\scriptsize Gain (ASPL)}
  & \includegraphics[width=\linewidth]{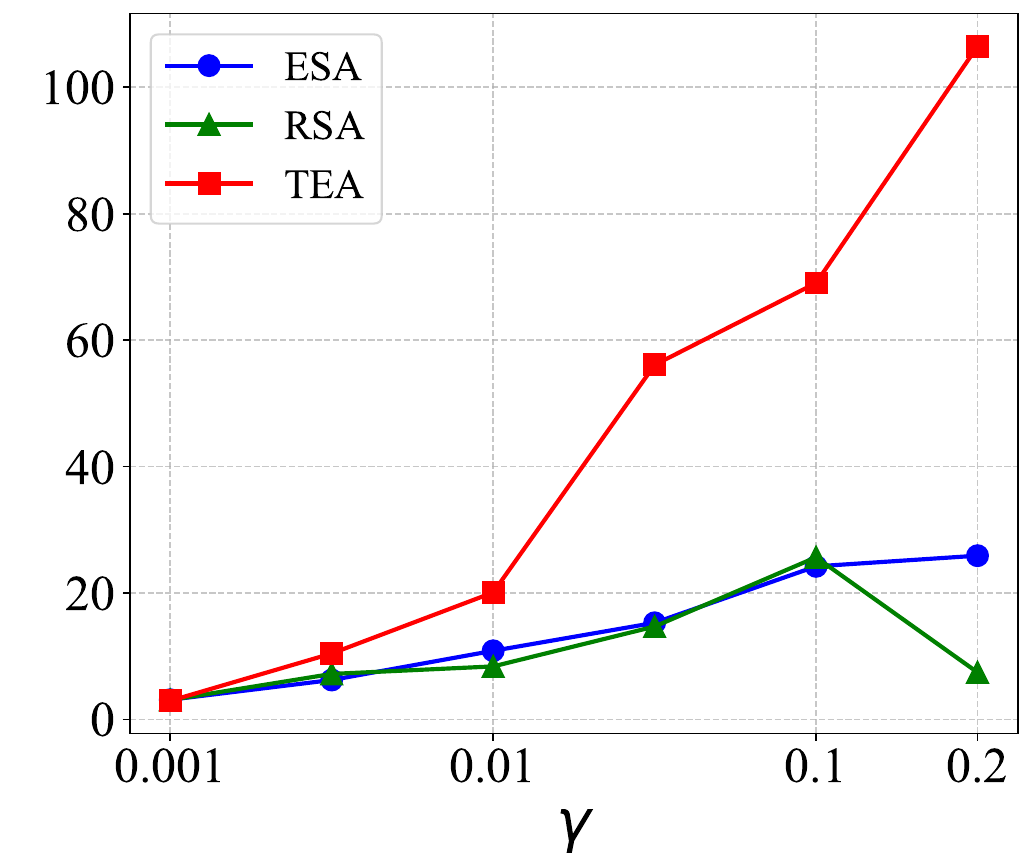}
  & \includegraphics[width=\linewidth]{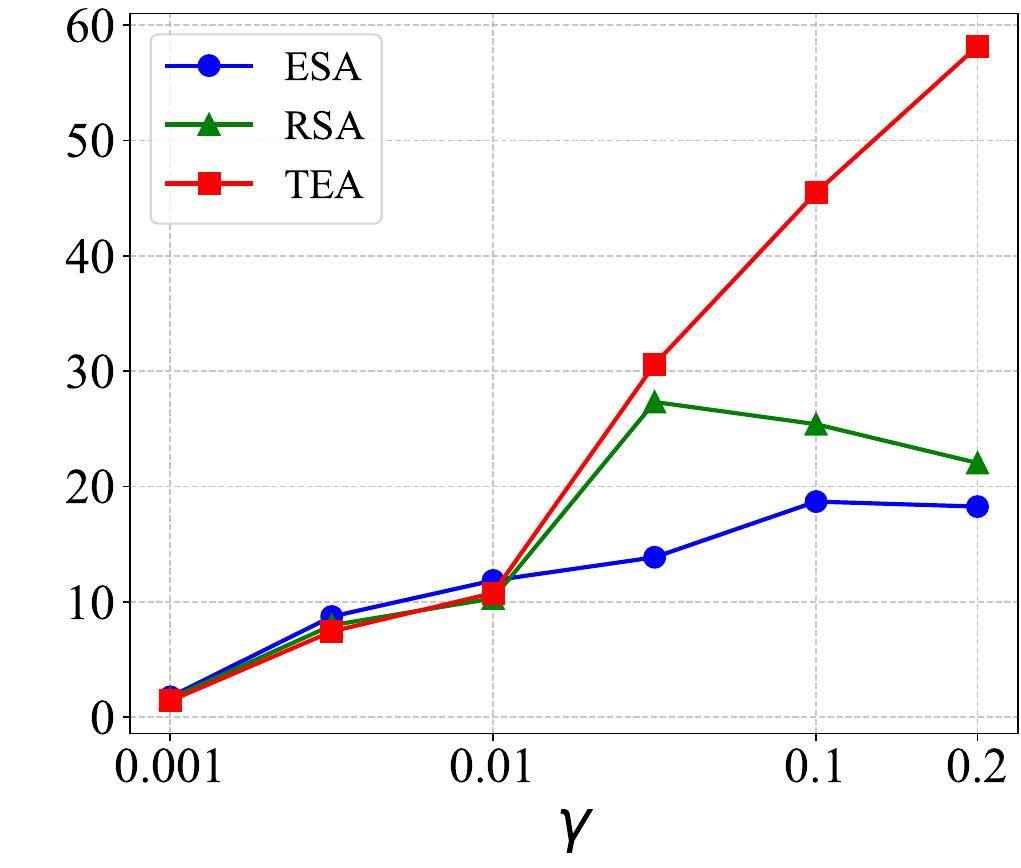}
  & \includegraphics[width=\linewidth]{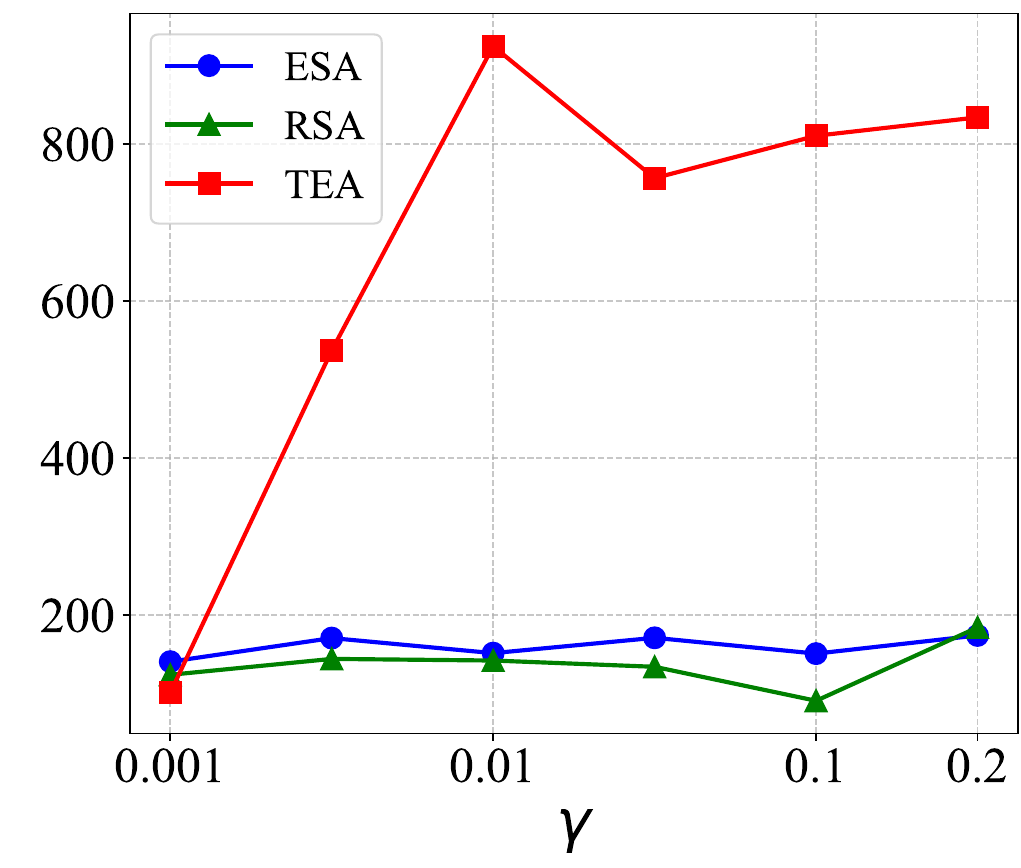}
  & \includegraphics[width=\linewidth]{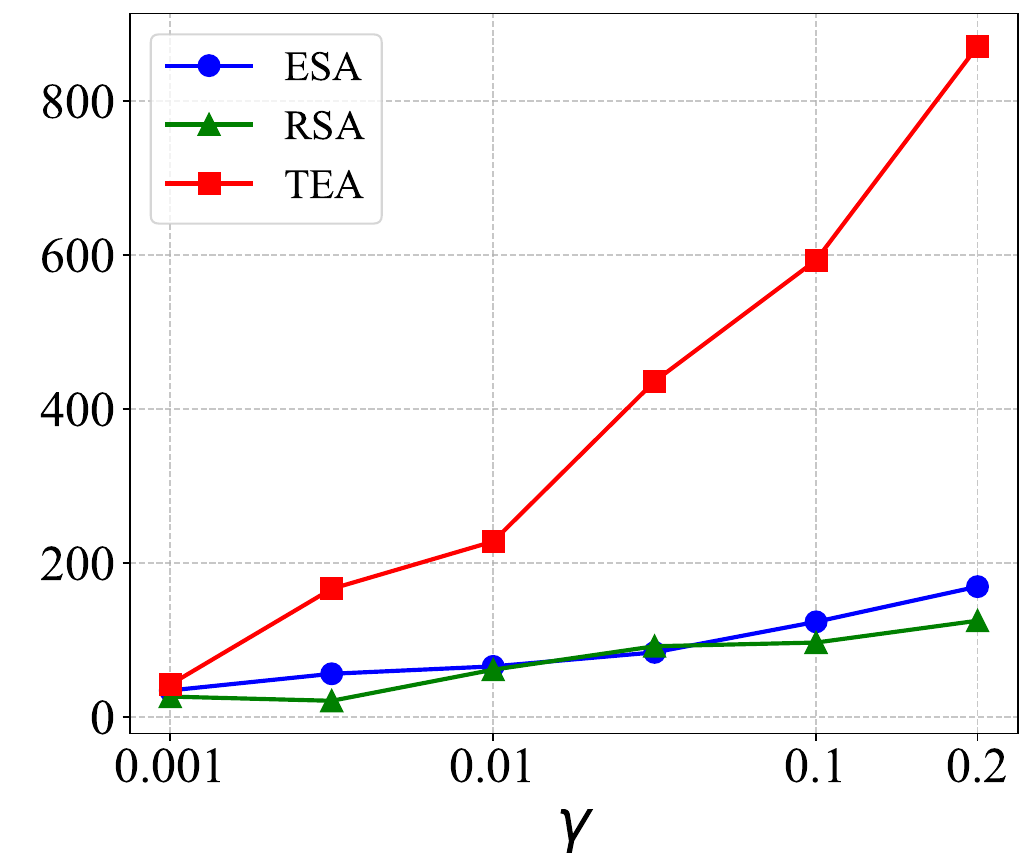} \\

\end{tabular}%
}
\caption{The effect of $\beta$, $\epsilon$ and $\gamma$ changes on gain of average shortest path length}

\label{fig:shortestpath_mga}
\end{figure*}

\textbf{Target node ratio $\gamma$:} The third row ($\gamma$ variation) shows an overall pattern in which, as $\gamma$ increases, TEA achieves a steadily increasing gain with the largest magnitude, ESA generally exhibits a gradual increase or remains stable, while RSA displays more pronounced fluctuations. This can be attributed to the expansion of the target node set with increasing $\gamma$: TEA benefits from concentrated budget allocation in high-weight time slices, which more effectively targets nodes and accumulates triangle perturbations; in contrast, RSA is affected by the random scheduling of attack scale, resulting in unstable coverage efficiency. Moreover, the increase of TEA is more pronounced in the Reply and Synthetic datasets, indicating that in such structural settings, target expansion has a stronger amplifying effect on triangle perturbations.

\subsection{Results of Shortest Path}

Figure \ref{fig:shortestpath_mga} illustrates the impact of varying parameters $\beta$, $\epsilon$, and $\gamma$ on the average shortest path gain. 
It should be noted that this metric is defined as the weighted sum of absolute changes in the shortest paths of target nodes. Therefore, a larger value indicates stronger perturbation to the path structure, rather than a monotonic increase or decrease in path length.

\textbf{Fake node ratio $\beta$:} As $\beta$ increases, all four datasets exhibit an overall upward trend, indicating that with a larger attack budget, both the number of injectable edges and reconstructable paths increase, thereby enhancing ASPL perturbation. TEA shows the fastest growth, with a clear acceleration in the medium-to-high $\beta$ range. Particularly, TEA achieves the most significant increase in Reply graph, suggesting a higher sensitivity to critical bridging edges.

\textbf{Privacy budget $\epsilon$:} As $\varepsilon$ increases, an overall “decrease–then stabilization or rebound” pattern is observed: in the low-to-moderate $\varepsilon$ regime, gains generally decline, indicating a compression of the effective attack space; in the mid-to-high $\varepsilon$ range, different datasets exhibit stabilization or rebound, reflecting a threshold effect. Notably, the Reply dataset shows the most pronounced rebound in TEA, suggesting the emergence of high-gain edge opportunities within this range; in contrast, Ass-733 stabilizes after the initial decline, indicating limited remaining perturbation capacity.

\textbf{Target node ratio $\gamma$:} Overall, as $\gamma$ increases, TEA achieves the most significant gain, ESA generally exhibits a gradual increase, while RSA shows notable fluctuations. This is primarily because the expansion of the target set introduces more perturbable paths, and under the same budget, TEA is more likely to target critical nodes in high-weight time slices. The Reply dataset exhibits the most pronounced fluctuations, indicating high sensitivity to target node combinations; in contrast, TEA and ESA in the Synthetic dataset show a steady upward trend, reflecting a more stable structural response.

Overall, parameter variations exert a certain regulatory effect on the results, while the intrinsic performance differences among methods remain the dominant factor influencing degree centrality gain. In addition, the attack effectiveness is closely related to the average degree of the dataset: a higher average degree allows fake nodes to establish more connections, thereby yielding more significant gains. For example, the Email dataset has the highest average degree and thus achieves the most pronounced improvement under the same parameter settings.

\section{Conclusion}
\label{sec:conclusion}
In this work, we conduct a systematic study of data poisoning attacks against privacy-preserving dynamic graphs under the local differential privacy (LDP) framework, and further explore their potential security implications. To capture the temporal characteristics of dynamic graphs, we introduce a novel attack effectiveness metric, namely absolute gain, which quantifies the cumulative impact of attacks on graph statistical metrics over time. Based on this metric, we design a temporal enhanced attack (TEA) strategy that adaptively schedules attack resources across time steps. We also develop two baseline strategies, equal-scale attack (ESA) and random-scale attack (RSA), to enable comprehensive evaluation and comparison.
Extensive experimental results on both real-world and synthetic graphs demonstrate that the proposed attack methods can significantly degrade the accuracy of graph metric estimation for target nodes. The findings reveal critical security vulnerabilities in existing LDP-based dynamic graph mechanisms. In particular, we observe that attackers can exploit temporal correlations in dynamic graphs to amplify attack impact, while current privacy-preserving approaches remain insufficiently robust against such adversarial behaviors.

For future work, it is important to design more robust LDP mechanisms and effective attack detection strategies to mitigate poisoning attacks in dynamic graph settings. Moreover, extending the proposed framework to a broader range of dynamic graph analysis tasks and evaluating its performance on larger-scale real-world graphs would further enhance the practical applicability and security of privacy-preserving dynamic graph systems.

\begin{acknowledgments}
The authors would like to thank the anonymous reviewers for their helpful comments. 
This work was supported by National Natural Science Foundation of China (62502519), Basic Research Program of Jiangsu (BK20251643).

During the preparation of this work, the authors used
ChatGPT to polish the language. After using this tool, the
authors reviewed and edited the content as needed and take
full responsibility for the final version of the manuscript.
\end{acknowledgments}





\bibliographystyle{IEEEtran}

\bibliography{refs}






\end{document}